\documentclass[aps,pra,superscriptaddress,twocolumn,tightenlines,showpacs,floatfix,amsmath,amssymb]{revtex4-1}

\usepackage{physics}
\usepackage{mathtools}
\usepackage{graphicx}
\usepackage{bm}
\usepackage{amssymb,amsmath,latexsym,mathrsfs}
\usepackage[usenames,dvipsnames]{color}
\usepackage[breaklinks,colorlinks,urlcolor=purple,citecolor=purple,linkcolor=purple]{hyperref}
\usepackage{float}
\usepackage{xcolor}
\usepackage{units}

\begin{document}

\title{Spontaneous irreversibility and objective thermalization in stochastic modifications of quantum theory}

\author{Aritro Mukherjee}
\thanks{\text{Corresponding author:  }aritro.mukherjee@uni-due.de; \newline aritro.m111@gmail.com.}
\affiliation{Faculty of Physics, University of Duisburg-Essen, Lotharstraße 1, 47048 Duisburg, Germany}
\affiliation{Institute for Theoretical Physics,
University of Amsterdam, Science Park 904, 1098 XH Amsterdam, The Netherlands}

\begin{abstract}
The deterministic and time-reversal symmetric dynamics of isolated quantum systems is at odds with irreversible equilibration observed in generic thermodynamic systems. Standard approaches at a reconciliation employ subjective restrictions on the space of observables or states and do not explain how a single macroscopic quantum system achieves equilibrium dynamically. We instead argue that quantum theory is an effective theory and requires corrections to accurately describe systems approaching the thermodynamic limit. We construct a stochastic extension of quantum theory which is practically identical to quantum mechanics for microscopic systems, yet allows individual macroscopic and isolated systems to objectively thermalize, generically. A fluctuation-dissipation relation guarantees physical consistency including norm preservation, energy conservation, no superluminal signalling and the emergence of microcanonical equilibrium. We further discuss the inclusion of objective collapse, thereby realizing a falsifiable theory of spontaneous universal irreversibility which describes the quantum-to-classical crossover dynamics of macroscopic quantum systems. The dynamics of the model describe spontaneous symmetry breaking, quantum state reduction and objective quantum thermalization for individual systems, while realizing emergent hybrid statistics for ensembles that integrate Born's rule and microcanonical equilibrium.\end{abstract}
\maketitle
\section{Introduction}\label{Sec1:Intro}
The irreversible approach of physical systems towards equilibrium is ubiquitous in nature and underlies the widespread success of equilibrium statistical mechanics at all length scales~\cite{PATHRIA2011,Stanf_Enc_phil24,Frigg2008_classica_therm}. However, how equilibrium is achieved dynamically, within closed quantum systems undergoing time-reversal symmetric and deterministic dynamics (such as via Schr\"odinger's equations) remains a foundational open question~\cite{ETHREview,Mori_2018_ETH_rev,ETH_typ_MBL_GGE}, which we term the quantum thermalization problem (QTP). 

\textcolor{black}{In the non-relativitic regime,} isolated quantum systems evolve unitarily via Schr\"odinger's equation which ensures that probability amplitudes do not change in the energy basis, up to a dynamical phase. This implies that the memory of the initial state is preserved and that \textcolor{black}{generically,} wavefunctions remains time dependent and cannot converge to any time-independent equilibrium state. Thus, although desirable, the strongest notion of an irreversible approach to equilibrium for an isolated system, one at the level of the (pure) quantum state, valid \textcolor{black}{without any restrictions on observables or initial states} of a given system, is trivially disallowed within standard quantum dynamics~\cite{ETHREview,Mori_2018_ETH_rev,ETH_typ_MBL_GGE}. 

Within the confines of standard quantum theory only various effective or weaker notions of irreversibility are tenable, which generally propose restrictions on the space of states or observables. These restrictions are motivated via epistemological arguments based on subjective considerations~\cite{aritroPhD,Jaynes_1957,Jaynes_1957_2} such as which observables are `viable' or `physical' in a given scenario  and proceed by prescribing a subset $\mathcal{A}_R \subset \mathcal{A}$ as relevant, out of all possible quantum observables in $\mathcal{A}$. Confined to this restricted algebra of observables, $\mathcal{A}_R$, such approaches are confined to ensemble averaged insights using coarse-grained mixed density operators (normal states on the subset $\mathcal{A}_R$ are generically mixed~\cite{Landsman2017}) which do not describe how a single system undergoes irreversible dynamics~\cite{Fortin2014,aritroPhD}. 

This is seen clearly by noting that for any quantum state $\hat{\rho}=\sum_{ij}\rho_{ij}\,\ket{i}\bra{j}$, if the chosen restricted algebra of observables, $\mathcal{A}_R\subset\mathcal{A}$, is such that for each observable $\hat{O}\in\mathcal{A}_R$, all transitions are disallowed, i.e. $\bra{i}\hat{O}\ket{j}=0\,\,(\forall\,i\neq j)$, then observable expectation values ($\forall \,\hat{O}\in\mathcal{A}_R$) are equal for both states, $\hat{\rho}$ and its (mixed) diagonal counterpart $\hat{\rho}'=\sum_i\rho_{ii}\,\ket{i}\bra{i}$. Clearly, the restricted set of observables $\mathcal{A}_R$ cannot distinguish between $\hat{\rho}$ and $\hat{\rho}'$ since these observables are incapable of resolving the off diagonal elements of $\hat{\rho}$. Other restrictions imposed on $\mathcal{A}$ lead to the equivalence of expectation values between different mixed states, which further, do not describe single systems~\cite{Fortin2014,aritroPhD}.

Usually, these restrictions are argued on the basis of epistemic motivations, such as which observables may or may not be accessible or reliably implemented by a given observer or physicist in a given scenario, henceforth termed an \textit{agent}. Thus, different restricted sets, $\mathcal{A}_R$ and $\mathcal{A}'_R$, may be seen as viable choices of observables for different agents. Since physical systems evolve independent of what is practically knowable, such restricted sets of observable algebras cannot be utilized to construct any objective notion of irreversibility or equilibration, a notion which should remain valid, \textcolor{black}{independent of any restriction on observables} and hence \textcolor{black}{valid for} all possible agents investigating a given quantum system~\cite{aritroPhD,Jaynes_1957,Jaynes_1957_2}. \textcolor{black}{It is in this sense we define objectivity, while subjective statements necessarily impose ignorance assumptions (on particular agents) such as by imposing restrictions on observables or on the discernibility of certain states.} As argued above, such constructions also necessarily ignore single system dynamics and is unable to describe or explain the dynamical emergence of irreversibility in a single, isolated quantum system, in the thermodynamic limit. Finally, within the confines of the axioms of standard quantum theory, the question of whether a given observable and quantum system thermalizes or not, has been argued to be (G\"odel) undecidable, impeding any axiomatic proof of such statements~\cite{Undecidibility2021}.

Motivated by these issues, in this article  we take an alternative view, that quantum theory is an effective theory and its dynamical equations require corrections or modifications to accurately describe systems with large number of degrees of freedom, i.e. for systems approaching the thermodynamic limit. We will show that in implementing this viewpoint with physical consistency, we recover not only an objective description of thermalization and single system irreversibility, but also, the dynamical emergence of equilibrium statistical mechanics.

Our motivation further draws from objective collapse theories~\cite{Bassi_03_PhyRep,Bassi2013Review,BohmBub_66_RevModPhys, Gisin84, Ghirardi_1986, Diosi_87_PLA, Ghirardi_90_PRA, percival95, aritroPhD,aritro2,aritro1,aritro3,Wezel10,Wezel_2009,Romeral_2024}, which aim at resolving the quantum measurement problem via modifications of quantum theory. The irreversible and random phenomenon of quantum state reduction or wave-function collapse during measurements also cannot be accounted for, within the time-reversal symmetric Schr\"odinger equation, and constitutes the quantum measurement problem~\cite{Wigner63,leggett2005quantum,Bassi_03_PhyRep,aritroPhD,Landsman2017}. Objective collapse theories resolve this problem by introducing small modifications to Schr\"odinger's equation in such a way that the unitary time evolution of microscopic particles is unaffected in any noticeable way, while the effect of the modifications dominate the dynamics in the macroscopic regime and cause quantum superpositions of large objects to reduce to classical configurations~\cite{Bassi_03_PhyRep,Bassi2013Review,BohmBub_66_RevModPhys, Gisin84, Ghirardi_1986, Diosi_87_PLA, Ghirardi_90_PRA, percival95, aritroPhD,aritro2,aritro1,aritro3,Wezel10,Wezel_2009,Romeral_2024}. In this context, quantum systems approaching the thermodynamic limit are also understood to be in the quantum-to-classical crossover regime: a regime where classicality is expected to emerge from the underlying quantum theory of its constituents. 

A typical objective collapse theory, however, cannot lead to any universal notion of thermalization, since the information of the initial conditions remain partially preserved via the so called Martingale condition, allowing the emergence of Born's statistics~\cite{Bassi_03_PhyRep,aritro2,aritro3,aritroPhD} (also see Appendix.~\ref{App.1}). Hence, corrections leading to objective collapse and those corrections leading to thermalization must be treated separately~\cite{aritroPhD}, the latter being the major focus of this article. 

After a brief discussion of thermalization in classical and quantum systems in Sec.~\ref{Sec2:Lit_Rev}, in Sec.~\ref{Sec3:OQT} we consider the strongest possible notion of quantum thermalization, that at the level of pure states \textcolor{black}{of individual closed systems}, and we discuss the requirements of a modified quantum theory with objective quantum thermalization (OQT). A generic form of an OQT model is argued, and we demonstrate how a fluctuation-dissipation relation enforces physicality conditions, such as norm preservation and the absence of superluminal signalling. Ensembles of systems are shown to evolve via a quantum semi-group.

Further, in Sec.~\ref{Sec3a}, the ensemble expectations of the \textcolor{black}{conserved} energy and the long-time steady states of the OQT model are used to extract stringent constraints and a unique form of the OQT model is established, which allows systems to equilibrate to a microcanonical distribution, while entropy increases and converges to Boltzmann's thermodynamic entropy at equilibrium. Novel physical consequences of the OQT model and the resulting observable deviations from standard quantum theory are then discussed. In Sec.~\ref{Sec3B}, protocols attempting to signal faster than light, using the OQT model are explored and shown to be disallowed. 

In Sec.~\ref{Sec4:SUI}, we integrate the OQT model with the previously established objective collapse models which are shown to be complementary, resolving both the quantum measurement problem and the quantum thermalization problem within the same theory. We term these models of Spontaneous Universal Irreversibility (SUI) as they constitute a minimal modification of quantum dynamics allowing macroscopic systems to spontaneously exhibit irreversibility and stochasticity, while realizing both classically localized and equilibrated states in their dynamics.  

Focusing on a recently proposed objective collapse model, we show that the hybrid SUI model describes three well known spontaneous irreversible phenomena for thermodynamic quantum systems---quantum state reduction, spontaneous symmetry breaking and objective thermalization, within the same dynamics. We discuss future directions and conclude in Sec.~\ref{Sec5:Concl}. Our results open up new possibilities of observational tests of fundamental physics and we hope to motivate a critical re-analysis of quantum interpretations and the foundations of equilibrium statistical mechanics.
\section{Classical and quantum thermalization}\label{Sec2:Lit_Rev}
This section briefly summarizes how standard approaches towards describing irreversibility and the approach of thermodynamic systems to equilibrium, necessarily employ so-called \textit{epistemic restrictions}. These are ultimately restrictions on what can be reliably known by any agent or investigator. If the approach to equilibrium, as explained using the standard dynamical principles is to be understood as a law of nature, then such restrictions on knowledge must also hold in principle, a premise we reject. \textcolor{black}{This section reviews how epistemic restrictions are employed in justifying irreversibility and equilibration in both classical and quantum systems. These arguments are agent-specific or subjective; in quantum systems, they further imply that any description of how an individual system dynamically thermalizes is necessarily ignored.} 
\subsection{{Classical Considerations}}Already in the classical regime, epistemic restrictions are employed to explain how physical systems approach thermal equilibrium. An isolated classical system evolves under deterministic and time‑reversal symmetric Hamiltonian dynamics~\cite{Arnold_CM1989}, which by itself, cannot yield an irreversible approach to a time-independent equilibrium state. Expressed differently, Hamiltonian dynamics ensures trajectories in phase space do not intersect (since they are one-to-one deterministic) implying that there can be no sinks or sources~\cite{Arnold_CM1989}, hence any irreversible convergence of trajectories to a time independent equilibrium state is not allowed generically. Thus some manner of epistemic restriction, such as coarse-graining or a restriction of observables is necessary. Moreover, spatially bound systems exhibit Poincaré recurrences, returning arbitrarily close to its initial state infinitely many times~\cite{Stanf_Enc_phil24,Frigg2008_classica_therm,Sklar1993Physics,Prigogine1980}. Both these issues, the problem of reversibility and the problem of recurrences~\cite{Stanf_Enc_phil24,Sklar1993Physics}, reappear in the quantum regime for (spatially bounded) isolated systems~\cite{ETHREview,Mori_2018_ETH_rev,ETH_typ_MBL_GGE}. 

\textcolor{black}{Classical equilibration stands on three pillars: Gibb's ensemble approach, Boltzmann's single system approach and arguments of typicality}~\cite{Stanf_Enc_phil24}. These approaches rely heavily on long-time averages and coarse-graining the phase (state) space to allow (mixed) densities instead of (pure) delta measures and also, coarse-graining the space of observables, motivated by what may be reliably known in specific scenarios, to specific agents~\cite{Stanf_Enc_phil24,Earman1996-EARWET,Wallace_classica_therm,Frigg2008_classica_therm,Mori_2018_ETH_rev,ETHREview}. 

In the Gibbs ensemble approach, coarse (mixed) probability densities (instead of delta measures) are employed, implying a lack of knowledge of the system's finer details or an uncertainty in the initial conditions. This is augmented with further coarse-graining of the phase space cells itself, restricting the investigation to macroscopic observables---smooth functions on the coarse phase space. \textcolor{black}{Ensemble averages of these observables are then considered and one tries to justify that they are time independent and coincide with measured values, which are argued to be temporal averages.}

Thus, further, one must argue some manner of mixing or ergodicity \textcolor{black}{on the energy shell}~\cite{Earman1996-EARWET,Stanf_Enc_phil24,Frigg2008_classica_therm} such that these long-time averaged quantities approach phase space ensemble averages under an appropriately coarse-grained thermal probability density or time invariant measure. However, physical systems where equilibration is expected, have not been definitively proven to possess constrained conditions such as ergodicity or more stronger properties like mixing~\cite{Earman1996-EARWET,Ergodic_heirachy_KAM_MM,Stanf_Enc_phil24}. We note that ergodicity by itself does not guarantee an approach to equilibrium nor an equilibration time, thus
ergodicity is not sufficient, but mixing is a sufficient criterion towards equilibration~\cite{Krylov_Mixing_nec}.  In specific situations of thermodynamic, separable systems with short range interactions and typical initial states, as well as crucially, for a specific class of macroscopic observables (sum functions), one may invoke Khinchin's notion of ergodicity ~\cite{Khinchin1949_1,Khinchin2} which does not remain valid otherwise.

\textcolor{black}{In the more general case, ergodicity is not generic. In integrable Hamiltonian systems, their phase spaces are foliated by invariant tori and each trajectory is confined to a torus determined by its initial conditions, leading to quasi-periodic dynamics without ergodicity \textcolor{black}{on the energy shell}~\cite{Arnold_CM1989}. Indeed, contrary to widespread belief, \textcolor{black}{partly} propagated by influential physicists like Fermi~\cite{Fermi1923_ergodic_wrong,CFLV_Castiglione_Falcioni_Lesne_Vulpiani_2008}, non-integrability does not automatically imply mixing or ergodicity. The Kolmogorov-Arnold-Moser (KAM) theorem~\cite{KAM_Arnold1963,KAM_Kolmogorov1954,KAM_Moser1962,Broer2004_KAM} shows that in near-integrable systems, a large-measure set of non-resonant invariant tori may deform but continue to persist and the system remains non-ergodic. Such systems exhibit Fermi-Ulam-Pasta-Tsingou recurrences resulting in non-ergodic behavior~\cite{FORD1992271_FPU_1}. For systems with three or more degrees of freedom, the complement of the surviving KAM tori may still support so-called Arnold diffusion~\cite{Arnold_diffusion}, but with exponentially slow time scales as estimated by Nekhoroshev’s theorem~\cite{Nekhoroshev1971,Nekhoroshev_Poschel1993}.}

Away from integrability, a more general result is the Markus-Meyer (MM) theorem~\cite{MM_MarkusMeyer1974,Ergodic_heirachy_KAM_MM}, which shows that ergodicity is not a generic property of (smooth) Hamiltonian systems and thus classical systems are not expected to equilibrate generically. It is important to note that MM considers only smooth Hamiltonian systems and thus, idealized situations such as hard ball gasses interacting via (non-differentiable) elastic collisions and Sinai billiards systems, which are known to be ergodic (for $N\ge2$ hard balls, see Ref.~\cite{Hard_ball_ergodicity,Hard_ball_2_Simanyi2004} and references therein for details) are not within its purview. This strongly suggests that barring such idealizations, in physically relevant systems, such as in systems with long range interactions, ergodicity is not generic~\cite{Non_Ergodic_LRO_1,Non-ergodic_LRO_2,Earman1996-EARWET}.

Indeed, mixing or ergodicity, with an appropriate coarse-grained measure, is usually justified by focusing on particular systems, most notably classical chaotic systems which showcase exponentially divergent trajectories of nearly identical initial conditions~\cite{Stanf_Enc_phil24,Frigg2008_classica_therm}. \textcolor{black}{However, chaos, understood as sensitive dependence on initial conditions, is a local property and does not necessarily imply mixing in the phase space globally. Global structures such as phase-space barriers can prevent trajectories from exploring the full energy surface, leading to non-ergodicity. In general, Hamiltonian systems possess mixed phase spaces divided into both regular non-ergodic as well as chaotic domains, with sticky borders where systems can persist for long times~\cite{Sticky_Hamiltonian,Sticky_ZASLAVSKY2002461,sticky_3}.}

Thus the above notions, although, allows ensembles of systems to approach equilibrium, however, they are not valid generically for all thermodynamic systems, all its initial states, or for all possible observables of a given system, and hence, yields a picture of equilibration based on epistemic restrictions. Thus, as argued before, different agents with differing coarse-graining of the phase space and possessing different coarse observables may not agree on what constitutes equilibrium. Indeed, the climax of this line of reasoning culminates in the maximum entropy interpretation of statistical mechanics and thermodynamics due to Jaynes~\cite{Jaynes_1957,Jaynes_1957_2}, wherein equilibrium is understood to be the maximal-entropy state consistent with the available, subjective information accessible to an agent. Said differently, it represents the best, consistent, state of knowledge and is maximally non-committal with regard to missing information. In this extreme view, neither equilibrium nor entropy, which is understood to be a lack of information, are objective, physical or dynamical properties of a system in and of itself~\cite{Jaynes_1957,Jaynes_1957_2}.

Focusing on the setting of single systems, Boltzmann provided the first microscopic derivation of the second law of thermodynamics, by establishing the celebrated H-theorem---showing the increase in thermodynamic entropy in a classical ideal (hard ball) gas, while making the so called \textit{Stosszahlansatz} or molecular chaos assumption, \textcolor{black}{that joint two-particle distributions always factorize as a product of two single particle marginal distributions}, an epistemic restriction neglecting correlations before and after collisions~\cite{Stanf_Enc_phil24,Frigg2008_classica_therm,Wallace_classica_therm,Sklar1993Physics}. Indeed, following arguments by Loschmidt and Zermelo that the Stosszahlansatz assumption is not consistent with the time-reversal symmetric and deterministic Hamiltonian laws of motion, Boltzmann developed two connected ideas, arguing for the increase of entropy, the notion of typicality and that the universe started from a lower entropic state~\cite{Stanf_Enc_phil24,Frigg2008_classica_therm}, as  discussed below. 

\textcolor{black}{Note that the Stosszahlansatz is the crucial step in closing the Bogoliubov–Born–Green–Kirkwood–Yvon (BBGKY) hierarchy~\cite{Cercignani1997ManyParticle_BBGKY} and  deriving the Boltzmann equation. Subsequently, in the same system of a dilute hard-sphere gas, Lanford ~\cite{Lanford1975TimeEvolution} proved that if a stronger form of the molecular chaos assumption holds at initial times (special non-overlapping initial conditions wherein all k-particle marginals factorize), it persists for short time intervals in the Boltzmann-Grad limit (number of particles tends to infinity while the hard ball radius tends to zero such that collision rate remains finite). More recently, Deng, Hani, and Ma have extended this result to hold for long times in the Boltzmann-Grad limit, provided that the corresponding Boltzmann equation has a regular solution (see Ref.~\cite{Deng_Hani_MA_2025} and its bibliography for relevant details). Nevertheless, these developments do not explain why the initial state should already satisfy molecular chaos or how it comes about. One may interpret the use of such special, chaotic states, as a specification of a coarse-graining on the phase space, so as to neglect correlations; this automatically also implies imposing restrictions on observables which could discern such fine-grained molecular correlations. We note that the modified laws proposed in Sec.~\ref{Sec3a} and Sec.~\ref{Sec4:SUI}, may provide the necessary objective mechanism for the Stosszahlansatz to arise in physical mesoscopic systems dynamically, shedding light on this paradox.}

Boltzmann's typicality ~\cite{Stanf_Enc_phil24,Frigg2008_classica_therm,Goldstein2010_qtypicality,Goldstein2010_QT,Tasaki2016_Qtypicality} observes that in thermodynamic systems, the phase space may be decomposed into sectors, where the observable expectation values of each micro-state in the largest sectors converge to a thermal expectation, for certain `viable' macroscopic observables. \textcolor{black}{The phase space sectors is indeed determined by these chosen observables and  labelled by their typical average values, while the deviations from these averages control the coarse-graining and determine the volume of these sectors.} Typically, it is argued that systems spend the largest time traversing the overwhelmingly large \textit{equilibrium} sectors where these macroscopic observables have appropriate thermal expectation values. 

Note that such statements depend on the phase space coarse-graining and do not forbid a system from spending timescales as large as the age of the universe out of equilibrium, before entering an equilibrium sector~\cite{Goldstein2010_qtypicality,Goldstein2010_QT}. Further, the fact that such equilibrium sectors dominate the phase space, do not imply that non-equilibrium states do not exist; in fact they may be prepared in laboratories and in some sense, the dynamical world at large, including biological systems, are out of equilibrium~\cite{Stanf_Enc_phil24,Prigogine1980,Sklar1993Physics}. Statements on the size of the non-equilibrium sectors approaching measure zero in the thermodynamic limit, \textcolor{black}{are firstly, contingent on the choice of observables and coarse-graining,} and further do not imply a negligible contribution to the system's dynamics~\cite{Sticky_Hamiltonian,sticky_3,Sticky_ZASLAVSKY2002461}, nor does it automatically imply that non-equilibrium states do not exist~\cite{Sklar1993Physics,Stanf_Enc_phil24}. 

Finally, to account for the observed increase in entropy, Boltzmann argued that the universe started in a \textit{preferred} low entropy state, such that there are many more ways of reaching a higher entropic equilibrium state~\cite{Stanf_Enc_phil24,Frigg2008_classica_therm,Sklar1993Physics,DavidAlbert2000TimeAndChance}. \textcolor{black}{In all scenarios, this is a necessity for equilibration: that in the past, the system state must have been in an exceedingly improbable and atypical macrostate, with respect to an unspecified choice of observables and coarse-graining. Apart from being somewhat vague, how such atypical initial conditions naturally arise is not addressed; further it also presents a restriction on which initial states may be prepared and can dynamically equilibrate.}

Crucially, note that in each above scenario, coarse-graining is required and the expectation values of only a restricted set of macroscopic observables converge to thermal expectation values, and such considerations allow, at best, an agent-dependent notion of what seems to be at equilibrium. One agent restricted to certain energy and time scales, determining their coarse-graining and chosen macroscopic observables, may not agree with other agents with access to more fine-grained observables. For example, an agent might adopt a coarse-graining such that the system’s entire phase space is treated as a single averaged cell and they may classify the system as being in equilibrium, since its coarse-grained observable expectation values do not change in time. \textcolor{black}{This deliberately extreme example already shows how epistemic restrictions influence the identification of thermodynamic equilibrium and the situation considerably worsens when multiple agents with conflicting criteria of equilibrium assess multiple systems.} Thus, in the above approaches, the notion of entropy and its increase, the approach to equilibrium and ultimately the derived thermodynamics are (partly) attributable to the lack of knowledge of agents, and not (solely) a consequence of the physical properties of systems in and of itself.  

\subsection{\textbf{Quantum Conundrums}}The issues discussed in the previous section continue to persist in quantum theory and shape arguments used towards a resolution of the quantum thermalization problem (QTP) in isolated systems. \textcolor{black}{In contrast to the classical case, a quantum pure state need not be restricted to a single energy shell and may instead be a superposition of many energy eigenstates. Indeed, quantum dynamics conserves the energy expectation value as well as the weights associated with each energy eigenspace, necessitating more stringent restrictions on `viable' initial states, as discussed below. Further, quantum measurements, although irreversible, adhere to Born's rule and preserve the memory of the initial conditions, thus cannot realize a thermal equilibrium state (see also Sec.~\ref{Sec4:SUI} and Appendix.~\ref{App.1}).} 

Early works by von Neumann extending the notion of ergodicity and typicality to the quantum case, already observed the requirement of restricting the space of observables (allowing only macroscopic observables which commute) and the state space, in addition to other requirements such as no resonances ~\cite{Neumann1929_QT,vonNeumann2010_QT,Goldstein2010_QT}. Consequent works by the Austin-Brussels group, although motivated by similar goals as this article---to universally account for irreversibility---focused on particular systems and observables, employing a strict ensemble worldview and abandoning any treatment of single systems~\cite{Prigogine1980,PRIGOGINE1999_AB1,Misra1988,LOCKHART198647,BISHOP20041_AB0}. The Zubarev school also attempted to explain quantum equilibration by instead focusing on non-standard modifications of ensemble dynamics and the master equations~\cite{Morozov1998_Zubarev_AB2}, however it is now well known that non-standard, especially non-linear modifications of the master equations may result in non-positivity and superluminal signalling~\cite{Gisin:1989sx,Bassi2015,aritroFTL,aritroPhD}. Other standard approaches employ the quantum Boltzmann equations which focus on neglecting correlations (like the classical Boltzmann approach) and approximating the analogous quantum collision terms in the master equations~\cite{VACCHINI200971,Eu1998_AB5,Bonitz2016QuantumKineticTheory}. Note that these approaches employ epistemic restrictions on the space of observables and neglect single systems, a common caveat.  

Indeed, contemporary approaches towards a resolution of the QTP, such as the eigenstate thermalization hypothesis (ETH)\cite{ETH0,ETH1,ETHRigol2008,Mori_2018_ETH_rev,ETHREview,ETH_typ_MBL_GGE,ETH_typ_MBL_GGE}, coupled with insights from open quantum systems and the decoherence paradigm~\cite{Zurek2009,Schlosshauer_2005,Breur_Petr02}, both, prescribe a restricted set of observables as `physically accessible' and admit purely agent-dependent notions of irreversibility. Again, such notions are only valid for specific, non-uniquely decomposable ensembles (with epistemic restrictions) and cannot constitute an agent independent or objective notion of equilibrium (valid without imposing restriction on observables or knowledge), nor can it explain how a single quantum system undergoes irreversible dynamics. 

Decoherence~\cite{Zurek2009,Schlosshauer_2005,Breur_Petr02} based approaches, focus on ensembles of a sub-system, or equivalently a restriction of observables to only sub-system observables $\mathcal{A}_S \subset \mathcal{A}$. Normal states on these restricted set of observables ($\mathcal{A}_S$) are generically mixed sub-system density operators, corresponding to the marginalized state of the sub-system averaged over all possible environment states. In other words, one restricts to `accessible' observables of the form, $\ \mathcal{A}_S\ni\hat{O}=\hat{O}_S\,\otimes\hat{\mathbb{I}}_E $, where $S$ represents the sub-system under study and $E$ represents its complement, usually an inaccessible environment ($\hat{\mathbb{I}}$ is the identity operator). This automatically implies that agents may possess only the information contained in reduced sub-system states, $\hat{\rho}_S=\mathrm{Tr}_E[\hat{\rho}]$ ($\mathrm{Tr}_E[...]$ denotes a partial trace operation), since for all $\hat{O}\in \mathcal{A}_S$, one has $\langle O\rangle=\mathrm{Tr} [\hat{O}\hat{\rho}]=\mathrm{Tr}[\hat{O}_S\,\hat{\rho}_S]$. These sub-system density operators or the corresponding observables may be shown to exhibit randomness and irreversible behavior under various assumptions such as a Markov assumption and the assumption that the environment is already at equilibrium~\cite{Lindblad1976,GKS76,Breur_Petr02,DeRoeck2010_QBM}. 

However, reduced (coarse-grained and marginal) density operators are generically mixed (non-uniquely decomposable as pure states) and are confined to describing \textit{ensembles}, which trivially implies their inability to describe a \textit{single} instance of irreversible evolution. In other words, open system approaches such as decoherence do not explain how a single system can undergo irreversible and random dynamics~\cite{Fortin2014,Adler_2003,aritro2,aritroPhD}. Further, different choices of decompositions of the entire state space into a sub-system and its complement, may showcase very different properties such as the presence of correlations and persistent memory of the initial conditions, thus leading to various agents (with access to different sub-system observables) disagreeing on what constitutes irreversibility and equilibrium.

Recently, the eigenstate thermalization hypothesis (ETH) has been subject to extensive investigations. ETH considers only those observables viable or physical which, in the energy basis, possess smoothly varying diagonal elements, while off-diagonals are suppressed and scale inversely with the system size~\cite{ETH0,ETH1,Mori_2018_ETH_rev,ETHREview,ETH_typ_MBL_GGE,ETH_typ_MBL_GGE}. These preferred, physically viable, coarse-grained observables, $\mathcal{A}_E \subset \mathcal{A}$ may \textit{seem} thermalized, while other observables in $\mathcal{A}$ will trivially never thermalize. The expectation values of these ETH-observables ($\mathcal{A}_E$), for most states in the Hilbert space confined to a narrow interval of energy, approximately converge on thermalized expectations and more so if the observable diagonal elements do not appreciably change within the energy window of interest. 

\textcolor{black}{However, such statements, apart from already imposing restrictions on observables and initial conditions, must also utilize long-time averaging (further restriction on observables with finite temporal resolution) and discard transient oscillatory contributions such as recurrences, justified by notions of quantum chaos which are encoded in energy-level statistics and reflected in the spectral form factor~\cite{ETHREview,BGS_new,Haake2018_QChaos}. Moreover, the spectral form factor itself is not self-averaging in general~\cite{SFF_not_self_averaging}; rather, one must necessarily resort to ensemble averaging over random Hamiltonians of different systems, energy-window smoothing, or temporal coarse-graining procedures, which may all be interpreted as a restriction of observables, coarse-graining as well as averaging over an ensemble of different systems. Finally, the Hamiltonian of a given quantum many body system is always structured and not a random matrix~\cite{ETHREview}; consequently, no objective notion of thermalization valid for any particular, individual system is ever established.}

Notions of quantum typicality~\cite{Goldstein2010_qtypicality,Tasaki2016_Qtypicality,ETH_typ_MBL_GGE,Goldstein_cannon_MC}---that almost every pure state yields thermal expectations for coarse macroscopic observables---are further used to augment ETH, yet again do not realize an agent-independent notion of thermalization, one valid \textcolor{black}{without any restriction on the system observables or its initial state}. Further, restricting to certain coarse algebra of observables, imply, again, that normal states on these algebras are mixed~\cite{Landsman2017}, i.e. the admissible density matrices are mixed and non-uniquely decomposable, which describe only ensembles and thus how an individual quantum system in a pure micro-state undergoes an irreversible approach to equilibrium is not described objectively. Note, observables such as $\ket{E}\bra{E'}+\ket{E'}\bra{E} \,\in\mathcal{A}$ with significant off-diagonals in the energy basis ($\ket{E},\ket{E'}$ are energy eigenstates) are not ETH observables, and they remain coherent and do not equilibrate within standard quantum theory. It is always possible to construct such ETH-violating observables which are typically argued away as being un-physical or inaccessible (to specific agents). Thus, again, ETH does not constitute an agent-independent notion of equilibration. Further, it is well known than ETH is evaded in many systems such as in integrable systems~\cite{ETHREview,Mori_2018_ETH_rev}, systems with quantum scarring and Hilbert space fragmentation~\cite{Fragmentation_hilbert_space,Moudgalya_2022_QMBS} as well as many-body localization~\cite{MBL1,ETH_typ_MBL_GGE}.

Recently, it has been argued that given the initial conditions and observables, the question of whether any given Hamiltonian evolution yields thermalized expectations or not, is mappable onto a halting problem, and already for certain simple systems, has been shown to be (G\"odel) undecidable within the mathematical framework of quantum theory and its axioms~\cite{Undecidibility2021}. Finally, since it is widely accepted that the universe is quantum and it is well known that in general, quantum systems may lack a classical analogue~\cite{Landsman2017}, the aforementioned issues motivate our exploration of modified axioms of quantum theory as a possible resolution of the QTP first.


In summary, the various notions of quantum thermalization admitted within standard quantum theory are agent dependent notions (valid for only particular observables) and do not explain how a single quantum system can undergo irreversible dynamics. Thus our main goals are to construct a viable notion of single system irreversibility and an agent-independent notion of equilibration. We will show that acknowledging the possibility that quantum theory is effective and may require corrections beyond its interval of validity, i.e. for systems approaching the thermodynamic limit, allows us to recover a stringently constrained modified quantum theory, which achieves both these goals, while being experimentally falsifiable. Further, we will show that the properties of the resulting modified quantum theory may reconcile and support the various approximations (such as Boltzmann's Stosszahlansatz) employed in recovering irreversibility and equilibrium in the standard context, for both quantum systems in the thermodynamic limit as well as their corresponding classical counterparts.

\section{Objective quantum thermalization}\label{Sec3:OQT}
Effective theories describe physical phenomena within its domain of validity and require corrections to preserve predictive power beyond this regime~\cite{Brauner2024Effective_SSB,Wezel_SSBlecturenotes}. Unless one dogmatically assumes the validity of quantum theory upto all length scales, seeking corrections or modifications beyond its expected regimes of validity is naturally warranted, if not necessary. Following this view, in this article, we modify quantum dynamics so as to allow irreversibility and equilibration dynamics for thermodynamic systems and our approach is motivated by the structure of objective collapse theories~\cite{aritroPhD,Bassi_03_PhyRep,Bassi2013Review,aritro1,aritro2,aritro3,Wezel10,Wezel_2009}, which offer a suitable framework to bypass the aforementioned epistemic restrictions~\cite{aritroPhD,DavidAlbert94,DavidAlbert2000TimeAndChance}.  

Objective collapse theories modify the quantum dynamics such that they are practically indistinguishable from quantum theory for microscopic systems, while for macroscopic systems, they allow an objective notion of quantum state reduction or wave-function collapse, applicable for all observables of an isolated system~\cite{Bassi_03_PhyRep,Bassi2013Review,aritroPhD,aritro1,aritro2,aritro3,Wezel10,Wezel_2009}. Crucially, ensembles of isolated systems undergoing objective collapse dynamics, at long times, follow a dynamical map, $\hat{\rho}\to\,\hat{\rho}_\mathrm{B}$ where $\hat{\rho}_\mathrm{B}$ is a statistical distribution corresponding to Born's rule~\cite{Bassi_03_PhyRep,aritroPhD,aritro2,aritro3}. 

Analogously, we are specifically interested in a modified quantum theory which can admit a dynamical map for ensembles, corresponding to the strongest notion of thermalization for isolated systems, wherein the state itself thermalizes,
$\hat{\rho}\,  \to \hat{\chi}$,
where $\hat{\chi}$ is a thermalized density matrix corresponding to an equilibrium distribution attained by an ensemble of isolated systems, such as the microcanonical distribution. \textcolor{black}{Thus, an  ensemble of identically prepared isolated systems, initially the same pure state, dynamically approach the equilibrium state, $\hat{\chi}$.} This strongest notion of thermalization is trivially inadmissible under the deterministic, unitary dynamics of Schrödinger’s equation, as well as under measurement or objective-collapse dynamics since they preserve information about the initial state. We note that Refs.~\cite{DavidAlbert94,DavidAlbert2000TimeAndChance} suggest that a notion of thermalization without epistemic restrictions is possible within objective collapse models, however we argue (and show explicitly in Appendix.~\ref{App.1}) that thermodynamic equilibrium of a closed system is not possible within a physically viable objective collapse model, due to the Martingale condition~\cite{Bassi_03_PhyRep,aritro2,aritro3} and Born's rule, which preserves the information of the initial conditions (also see Ref.~\cite{Ag_DavidAlbert2021}).

Indeed, in contrast to previous approaches, our approach focuses on \textcolor{black}{stochastic} modifications of the dynamics of isolated quantum systems, allowing an objective resolution of the QTP, independent of details of (thermodynamic) systems, initial conditions or preferred observables. Further, we ensure that the modifications scale with the system size such that, for microscopic systems, the modified dynamics is practically indistinguishable from quantum theory, while macroscopic systems thermalize objectively. Due to the differing dynamics of macroscopic quantum systems from its standard expectations, such theories are in principle falsifiable in controlled experiments in the mesoscopic regime, similar to objective collapse theories~\cite{Bassi2013Review,Wezel_2012,Carlesso_2022,Romeral_2024}. We will now construct a model allowing objective thermalization for a macroscopic, isolated, quantum system with Hamiltonian $\hat{H}$. We consider a (not necessarily finite) Hilbert space, $\mathcal{H}$ with countable energy eigenstates, $\hat{H}\ket{\mu}=E_\mu\ket{\mu}$. 

To construct the model we first argue that unlike objective collapse models which possess the Martingale property~\cite{oksendal2003stochastic,Bassi_03_PhyRep,aritro2,aritro3}, any objective thermalization model cannot posses this Martingale property in the energy basis, since this would preserve the information of the initial conditions~\cite{aritroPhD} (also see Appendix.~\ref{App.1}). Thus, a viable starting point for the so-called thermalization operators in any modified quantum theory allowing an objective notion of thermalization, must present possible transitions between energy eigenstates (denoted $\ket{\mu}, \,\ket{\nu}$). 

The primitive family of operators allowing such transitions are $\hat{L}_{\mu\nu}=\ket{\mu}\bra{\nu}$, which determines the form of an un-normalized quantum (Ito) stochastic process~\cite{Bassi_unique,Gisin:1989sx,Bassi_03_PhyRep} on $\mathcal{H}$, given by ($\hbar=1$):
\begin{align}\label{Eq:therm1}
    & \,d\ket{\psi} =-i\hat{H}\ket{\psi}dt + \sum_{\mu,\nu} d\hat{G}^{\mu\nu}\ket{\psi},\\
    &d\hat{G}^{\mu\nu}\ket{\psi}:= \,\,D^{\mu\nu}\,\bigg[\hat{{L}}_{\mu\nu} -\langle\hat{L}_{\mu\nu}\rangle\bigg]\ket{\psi} \, dW_t^{\mu\nu} \nonumber \\&\,+ J^{\mu\nu}\,\, \bigg[\langle\hat{{L}}^\dagger_{\mu\nu}\rangle\hat{{L}}_{\mu\nu}-\frac{1}{2}\hat{{L}}^\dagger_{\mu\nu}\hat{{L}}_{\mu\nu} -\frac{1}{2}\langle\hat{L}_{\mu\nu}\rangle\langle\hat{{L}}^\dagger_{\mu\nu}\rangle\bigg]\ket{\psi}\, dt. \nonumber
\end{align}
Here, $\ket{\psi}$ is the time dependent wave function and $\hat{H}$ is the standard Hamiltonian, while $d\hat{G}^{\mu\nu}$ is the stochastic modification, to be constrained further. When the modification, $d\hat{G}^{\mu\nu}=0,\,\forall\,\mu,\,\nu,$ Eq.~\eqref{Eq:therm1} reduces to the standard Schr\"odinger equation evolving isolated systems. Here $\langle\hat{O}\rangle=\bra\psi\hat{O}\ket{\psi}$ is the usual (time dependent) quantum expectation value for an individual system in the state $\ket{\psi}$. 

To ensure that these modifications are significant for macroscopic quantum systems, the strength of the modifications must scale with the (effective) number of degrees of freedom of the system implying that $d\hat{G}^{\mu\nu}$ is extensive and proportional to a phenomenological coupling $\mathcal{N}$. Within each $d\hat{G}^{\mu\nu}$ there are two terms controlling the rate of transitions, one is non-linear and deterministic, with the coupling $J^{\mu\nu}$, while the other term is stochastic, coupled via $D^{\mu\nu}$. Both terms must be proportional to $\mathcal{N}$, as further clarified by the fluctuation dissipation relationship discussed below. Note that both coefficients are non-negative real-valued numbers, and are not symmetric, i.e. $J^{\mu\nu}\neq J^{\nu\mu}$ and $D^{\mu\nu}\neq D^{\nu\mu}$. 

The factor $dW^{\mu\nu}_t$ indicates real valued Gaussian increments of independent standard Wiener processes (i.e. corresponding to the Brownian motion $W_t=\int\,dW_t$ with $W_0=0$)~\cite{oksendal2003stochastic,gardiner2004handbook}. In the above expression note that $dW^{\mu\nu}_t$ and $dW^{\nu\mu}_t$ are increments of independent Wiener processes. We use the convention that each $dW^{\mu\nu}_t$ is sampled from a Gaussian distribution with standard deviation $\sqrt{dt}$ which imply the standard time independent expectation values, $\mathbb{E}\left[dW^{\mu\nu}_t\right] = \mathbb{E}\left[W^{\mu\nu}_t\right]=\mathbb{E}\left[dt \,dW^{\mu\nu}_t\right]=0$, $\mathbb{E}\left[dW^{\mu\nu}_t\,dW^{\mu\nu}_s\right]=0$ for $t\neq s$ and at equal times, $\mathbb{E}\left[dW^{\mu\nu}_t\,dW^{\mu'\nu'}_t\right]=\delta_{\mu\mu'}\delta_{\nu\nu'}dt$ or simply, $(dW^{\mu\nu}_t)^2=dt$ ~\cite{gardiner2004handbook, oksendal2003stochastic}. Here $\mathbb{E}[\,\cdot\,]$ indicates the average over an ensemble of realizations of the Wiener process and hence, trajectories of Eq.~\eqref{Eq:therm1}~\cite{gardiner2004handbook, oksendal2003stochastic,aritro2,aritroPhD}. Note that we assume that this stochastic influence arises from physics beyond quantum mechanics, and it is not an averaged description of the influence of unobserved quantum degrees of freedom~\cite{Bassi_03_PhyRep,Bassi2013Review,aritro1,aritro2,aritroPhD}.

We now consider constraints such that the modifications do not change the kinematic character of quantum theory. Particularly, norm-conservation is upheld in the dynamics of Eq.~\eqref{Eq:therm1} by observing a fluctuation dissipation relationship (FDR) between the stochastic and deterministic components of the modifications~\cite{aritro1,aritro2,aritro3,aritroPhD}. Concretely, using the norm, $N_\psi=\bra{\psi}\psi\rangle$ and applying Ito's lemma, $dN_\psi= \bra{d\psi}\psi\rangle + \bra{\psi}d\psi\rangle +\bra{d\psi}d\psi\rangle \,\,$, we find its change, $dN_\psi =0$ (at all times) with an FDR of the form $(D^{\mu\nu})^2\,=\,J^{\mu\nu}\, \propto \,\mathcal{N}$. This implies that within an ensemble of identically prepared systems, each ensemble member individually undergoes the stochastic and norm-conserved dynamics of Eq.~\eqref{Eq:therm1} while the FDR is upheld.

To find how the ensemble evolves we consider the evolution of the pure state projector, $\hat{\Psi}:= \ket{\psi}\bra{\psi}$, which describes how a single ensemble member evolves. Here the ensemble constitutes individual systems evolving via Eq.~\eqref{Eq:therm1} with differing stochastic trajectories. The stochastic average over these, $\mathbb{E}[\hat{\Psi}] = \hat{\rho}$, yields how the entire ensemble density evolves. To compute this, we use Ito's lemma~\cite{oksendal2003stochastic,gardiner2004handbook}, $d\hat{\Psi}=|d\psi\rangle\langle\psi|+|\psi\rangle\langle d\psi|+|d\psi\rangle\langle d\psi|$, and average over the stochastic components to obtain the evolution of the corresponding statistical operator or density matrix, with $\mathbb{E}[d\hat{\Psi}] = d\hat{\rho}$. This yields the linear master equations of the Gorini-Kossakowsky-Sudarshan-Lindblad (GKSL) form~\cite{Lindblad1976,GKS76,Breur_Petr02}, given by ($\hbar=1$):
\begin{align}&\frac{\partial\hat{\rho}}{\partial\,t} =
-i\left[ \hat{H} \,,\, \hat{\rho} \right] \nonumber\\&+\, \mathcal{N}\sum_{\mu,\nu} J^{\mu\nu}\bigg(  \hat{{L}}_{\mu\nu} \hat{\rho}\hat{{L}}^\dagger_{\mu\nu} - \frac{1}{2}\Big\{\hat{{L}}^\dagger_{\mu\nu}\hat{{L}}_{\mu\nu}\,,\,\hat{\rho}\Big\}\bigg).\label{Therm_master_0}\end{align}
Here $\hat{\rho}$ is the noise averaged statistical operator (density matrix) for an ensemble of (isolated) systems undergoing the dynamics of Eq.~\eqref{Therm_master_0} with the FDR, $(D^{\mu\nu})^2\,=\,J^{\mu\nu}$. The proportionality factor scaling with the system size, $\mathcal{N}$, has been extracted from the definition of the coupling constants. Note that the above master equation, unlike those obtained in objective collapse theories, resembling dephasing GKSL equations~\cite{Bassi_03_PhyRep,aritro2,aritro3}, takes a form explored in the context of open system equilibration via detailed balance~\cite{Gorini_Frgerio_verri_KS78, Houghston_therm}. The above equations allow us to constrain the model further so that an appropriate steady state is reached, given the system Hamiltonian. However, we stress here that in this model, a (linear) quantum semi-group is obtained for the dynamics of an ensemble of isolated systems and unlike other approaches, does not integrate out inaccessible parts of the setup. Instead, the master equations results from averaging over the modified stochastic quantum dynamics where each ensemble member is a single instance of an isolated system undergoing the dynamics of Eq.~\eqref{Eq:therm1} with the FDR.  

\subsection{\textbf{Energy, Entropy and Equilibrium}}\label{Sec3a}
Having obtained the ensemble dynamics of isolated quantum systems in Eq.~\eqref{Therm_master_0}, we will now show that constraints arising from its  equilibrium steady states and the average energy conditions, impose stringent constraints on the model and the form of $J^{\mu\nu}$. This will allow us to construct an energy conserving theory of objective quantum thermalization viable for isolated systems.

Our main goal is ensure that for any (non-equilibrium) state $\hat{\rho}$, the dynamics of Eq.~\eqref{Therm_master_0} guarantees that a steady equilibrium state is reached at long times. Denote these steady states corresponding to an equilibrium density (statistical) operator as $\hat{\chi}$, which is time independent, mixed and diagonal in the energy basis. $\hat{\chi}$ is diagonal since it represents an average over an ensemble and off-diagonal contributions are necessarily averaged out, as noted in both the quantum microcanonical ensemble and canonical ensemble, which are diagonal in the energy basis~\cite{PATHRIA2011,Mori_2018_ETH_rev, ETH_typ_MBL_GGE,ETHREview,Goldstein2010_qtypicality,Goldstein_cannon_MC}. We will obtain these steady states as solutions of the constraints below, however it is important to note that $\hat{\chi}$ ultimately represents the empirically observed equilibrium distribution, reached by an ensemble of identical, isolated (macroscopic) quantum systems.

Imposing the steady state condition, $\frac{\partial\hat{\rho}}{\partial t}=0$, in Eq.~\eqref{Therm_master_0}, we find the first (operator) constraint, $\hat{\mathcal{C}}_\chi=0$, towards achieving the equilibrium steady state $\hat{\chi}$, given by:
\begin{align}
\hat{\mathcal{C}}_\chi:=\sum_{\mu,\nu}J^{\mu\nu} \chi_\nu\bigg(  \,\ket{\mu}\bra{\mu} -  \ket{\nu}\bra{\nu}\bigg). \nonumber
\end{align}
Here, $\chi_\nu = \bra{\nu}\hat{\chi}\ket{\nu}$ and $\sum_\nu\chi_\nu=1$ since $\mathrm{Tr}[\hat{\chi}]=1$. Phrased differently, if the dynamics of Eq.~\eqref{Therm_master_0} leads to a long-time equilibrium steady state of the form $\hat{\chi}$, then the constraint $\hat{\mathcal{C}}_\chi=0$ must be structurally satisfied by the corresponding Lindblad form. 

Consider now the average energy, $E=\mathrm{Tr}[\hat{\rho}\,\hat{H}]$, of the ensemble  of isolated systems, using Eq.~\eqref{Therm_master_0} and the (standard) Hamiltonian, $\hat{H}=\sum_\mu E_\mu\ket{\mu}\bra{\mu}$. The constraint ensuring energy conservation is  $\mathcal{C}_{E}(t)=0$ where $\mathcal{C}_{E}(t):=dE/dt$. Note, the average energy is conserved if and only if the constraint, $\mathcal{C}_{E}(t)=0$ is satisfied at all times, and for all states on the Hilbert space. Without assuming any restriction on $\hat{\rho}$, this constraint takes the form:
\begin{align}
    \mathcal{C}_{E}(t):=\sum_{\mu,\nu}J^{\mu\nu}\,\rho_{\nu\nu}\,\big(  E_\mu\, -  E_\nu\,\big). \nonumber
\end{align}
Here, $\rho_{\nu\nu}=\bra{\nu}\hat{\rho}\ket{\nu}$ is time dependent and in general, may be any (normalized) state. Since any physically viable, objective thermalization model cannot allow for unrestricted changes in ensemble averaged energies, but must yield a long-time equilibrium state, one must solve for both $\hat{\mathcal{C}}_\chi=0$ and $\mathcal{C}_{E}(t)=0$ together and obtain a form of $J^{\mu\nu}$. 

To do this, first, we use an ansatz, $J^{\mu\nu}=A^\mu B^\nu$ in $\hat{\mathcal{C}}_\chi=0$ and $\mathcal{C}_{E}(t)=0$, which yield the following two suggestive expressions respectively:
\begin{align}
\hat{C}_\chi:=&\,\,\,\sum_{\mu}A^\mu\sum_{\nu}\, \chi_\nu\,B^\nu\bigg(  \,\ket{\mu}\bra{\mu} -  \ket{\nu}\bra{\nu}\bigg)=0, \nonumber\\
\mathcal{C}_{E}(t):=&\,\,\,\sum_{\mu}A^\mu\sum_{\nu}B^\nu \big(  E_\mu\, -  E_\nu\,\big)\rho_{\nu\nu}=0. \nonumber
\end{align}
It is seen that both conditions may be uniquely met by the asymmetric choice of $B^\nu=1$ yielding ${A^\mu}{/\mathcal{Z}} =\chi_\mu$ with $\sum_\mu A^\mu =\mathcal{Z}$, observed from the promising equalities:
\begin{align}
\hat{C}_\chi:=\,\,\,\,\,\,&\, \sum_{\mu}\bigg(\frac{A^\mu}{\mathcal{Z}} - \chi_\mu\bigg)\,  \,\ket{\mu}\bra{\mu} =0, \label{Eq:C1_fin}\\
\mathcal{C}_{E}(t):=\,\,\,\,\,\,&\sum_{\mu}E_\mu \frac{\, A^\mu}{\mathcal{Z} } \, =  {\sum_{\nu}\, E_\nu\,\rho_{\nu\nu}. \,}\label{Eq:C2_fin}
\end{align}
The first equality encodes the requirement that the Hilbert space of the quantum system must possess a time invariant thermal state $\hat{\chi}$ and is related to the quantum semi-group in Eq.~\eqref{Therm_master_0} via $J^{\mu\nu}=A^\mu=\chi_\mu\,$ ( $\forall\,\mu,\nu$ with $\mathcal{Z}=1$). The second equality ensures energy conservation in the ensembles, and shows again that the $A^\mu$ are weights of a probability distribution. Note that while the right hand side is time dependent and equals $\mathrm{Tr[\hat{\rho}\hat{H}]}$, the left hand side is independent of time, with ${\sum_{\mu}E_\mu \, A^\mu} = \mathrm{Tr}[\hat{\chi}\hat{H}]$, using Eq.~\eqref{Eq:C1_fin}. Then, if the condition in Eq.~\eqref{Eq:C2_fin} holds at initial times, it will hold for all times since Eq~\eqref{Therm_master_0} is a quantum semi-group. 


However, multiple solutions of $\hat\chi$ are possible which satisfy the above constraints. This is not unexpected given the fact that in a differing setting of (standard) open quantum systems, Eq.~\eqref{Therm_master_0} may be interpreted as the reduced dynamics of a sub-system after averaging out an inaccessible environment. Thus, in the scenario of a single conserved quantity which is the energy, $\hat\chi$ may be associated with a canonical distribution, $\hat{\chi}_\beta= e^{-\beta\hat{H}}/\mathcal{Z}$, where $\mathcal{Z}=\mathrm{Tr}[e^{-\beta\hat{H}}]$ is the partition function and $\beta$ is set appropriately at initial times. Note that $\hat{\chi}_\beta$ satisfies the above constraints and we can always expect such a distribution to exist in a macroscopic system. However, since we are interested in isolated systems, we instead focus on the quantum microcanonical distribution denoted $\hat{\chi}_E$, which is the expected long-time equilibrium distribution for an ensemble of isolated, macroscopic quantum systems~\cite{PATHRIA2011}. 

The quantum microcanonical distribution~\cite{PATHRIA2011, Goldstein_cannon_MC,Goldstein2010_qtypicality} for an ensemble of identical quantum systems with energies between $E$ and $E\pm\delta E$, is given by an equally weighted sum, $\hat{\chi}_E= \frac{1}{\Omega}\sum_\mu \hat{\mathbb{P}}_\mu$ where the sum is over all energy eigenstates, $\hat{\mathbb{P}}_\mu=\ket{\mu}\bra{\mu}$ which possess energies $\mathrm{Tr}[ \hat{H}\,\hat{\mathbb{P}}_\mu]\in[E-\delta E,E+\delta E]\,\,\forall \mu$. Here $\Omega$ is the number of such micro-states such that $\mathrm{Tr}[\hat{\chi}]=1$. Further we expect for any macroscopic quantum system, such a $\hat{\chi}_E$ exists and satisfies Eq.~\eqref{Eq:C2_fin}, for all possible quantum states, $\hat\rho$, with the same energy, $\mathrm{Tr}[\hat\rho \hat{H}]=\mathrm{Tr}[\hat{\chi}_E \hat{H}]$. 

However, we still need to determine $\delta E$, which must be derived from the initial conditions of the (isolated) system. To do this, we focus on the evolution of the variance of the energy, $V= \langle\hat{H}^2\rangle -\langle\hat{H}\rangle^2$. Using Eq.~\eqref{Therm_master_0}, $J^{\mu\nu}=\chi_\mu$ and Eq.~\eqref{Eq:C2_fin} we find:
\begin{align}
\dot{V} = \sum_\mu E_\mu^2 \, \chi_\mu \,- \sum_\nu E_\nu^2 \, \rho_{\nu\nu}.\nonumber
\end{align}
The above expression shows that the change in the variance of the ensemble's energy depends on the difference between the variance of the steady state, $\hat{\chi}$ and the state undergoing the evolution, $\hat{\rho}$. With the choice of the steady states, corresponding to the canonical distribution $\hat{\chi}_\beta$, it is seen that the variance of the energy will always evolve, for any initial (non-equilibrium) state and converge to that of $\hat{\chi}_\beta$. 

However, with the choice of $\hat{\chi}_E$, the microcanonical ensemble, if we further identify $|\delta E|=\sqrt{V}$, then we fix the conditions such that the average energy variance also does not change. Here, we thus make an assumption, which must ultimately be tested in experiments. We fix the model such that at long times, a macroscopic quantum system known to be in a non-equilibrium state with energy $E$ and variance $V=\delta E^2$, approaches a quantum microcanonical distribution $\hat{\chi}_E$, which is diagonal and uniformly distributed at all energy eigenstates lying in the microcanonical window of $[E-\delta E,E+\delta E]$. This agrees with the conventional expectations for thermodynamic limit quantum systems and the standard definitions of the quantum microcanonical ensemble~\cite{PATHRIA2011,Goldstein2010_qtypicality, Goldstein_cannon_MC}. 

We further note that although the choice of $\hat{\chi}_\beta$ does not keep energy eigenstates (except the ground state) stable under evolution, the choice of $\hat{\chi}_E$ using the above prescription does indeed keep all energy eigenstates stable. Concretely, if the initial state is a non-degenerate energy eigenstate, $\ket{\phi}=\ket{\gamma}$, the choice of $\hat{\chi}_E=\ket{\gamma}\bra{\gamma}$ is fixed by its zero variance and preserves the initial state under the action of Eq.~\eqref{Therm_master_0}, with the above constraints. Finally, since we expect closed systems to evolve to a microcanonical distribution, the choice of $\hat{\chi}_E$ is ultimately more preferable and ensures a consistent notion of objective quantum thermalization for isolated systems. 

Note that the above procedure may be generalized to settings with multiple conserved quantities. While in the case of the canonical solutions $\hat{\chi}_\beta$, these would correspond to the generalized Gibbs ensemble~\cite{ETH_typ_MBL_GGE}, in the case of microcanonical solutions, $\hat{\chi}_E$, one simply ensures that conserved quantities remain preserved. Thus for each symmetry generator, $\hat{S}$ of the Hamiltonian, we must ensure that $\frac{\partial}{\partial t}\langle \hat{S}\rangle =0 $, establishing further constraints. 

Using Eq.~\eqref{Therm_master_0}, one scenario in which these constraints are satisfied is if $\hat{S}$ commutes with $\hat{L}_{\mu\nu}$, showing that transitions are allowed only for micro-states within the sectors fixed by the initial value of the conserved quantity, denoted $S=\langle \hat{S}\rangle$ and its variance $\delta S^2 = \langle \hat{S}^2\rangle-\langle \hat{S}\rangle^2$, like in the previous case. Then in this scenario, the microcanonical distribution is given by, $\hat{\chi}_{(E,S)} = \frac{1}{\Omega} \sum_\psi \hat{\mathbb{P}}_\psi$ where $\hat{\mathbb{P}}_\psi=\ket{\psi}\bra{\psi}$ are projections onto the energy basis, and the sum is over all energy eigenstates, such that  $ \langle\psi |\hat{H}|\psi\rangle\in[E-\delta{E}, E+\delta{E}]$  and further, $\langle\psi |\hat{S}|\psi\rangle\in [S-\delta S,S+\delta S]$ for each micro-state. As before, $\Omega$ is the number of such micro-states such that $\mathrm{Tr}[\hat{\chi}]=1$. A more detailed analysis of the constraints pertaining to multiple conserved quantities is left for the future. For our treatment, it suffices that a microcanonical $\hat\chi$ exists for any given (thermodynamic) system. 

With these constraints satisfied, we are now in a position to write down the main expressions for the modified quantum dynamics, allowing an objective notion of thermalization, with stable end states as an appropriate quantum microcanonical distribution, $\hat{\chi}_E$ with energy as its conserved quantity. Individual macroscopic quantum systems with (initial) quantum energy expectation, $E_\psi=\bra{\psi}\hat{H}\ket{\psi}$ and variance $V_\psi= \langle\hat{H}^2\rangle -\langle\hat{H}\rangle^2$ evolve as per the following (norm preserving) quantum stochastic process:
\begin{align}\label{Eq:thermFinal}
    &\,d\ket{\psi} =-\frac{i}{\hbar}\hat{H}\ket{\psi}dt + \sum_{\mu,\nu} d\hat{G}_\chi^{\mu\nu}\ket{\psi},\\
    &d\hat{G}_\chi^{\mu\nu}\ket{\psi}:= \,\,\sqrt{\mathcal{A}^\mu}\,\bigg[\hat{{L}}_{\mu\nu} -\langle\hat{L}_{\mu\nu}\rangle\bigg]\ket{\psi} \, dW_t^{\mu\nu}  \nonumber\\&+\mathcal{A}^\mu\,\, \bigg[\langle\hat{{L}}^\dagger_{\mu\nu}\rangle\hat{{L}}_{\mu\nu}-\frac{1}{2}\hat{{L}}^\dagger_{\mu\nu}\hat{{L}}_{\mu\nu} -\frac{1}{2}\langle\hat{L}_{\mu\nu}\rangle\langle\hat{{L}}^\dagger_{\mu\nu}\rangle\bigg]\ket{\psi}\, dt.\nonumber 
\end{align}
Here, $\mathcal{A}^\mu=\big(\,{\alpha\,\mathcal{N}}/{\hbar}\,\big)\,\chi^\mu_E$ where $ \chi^\mu_E=\bra{\mu}\hat{\chi}_E\ket{\mu}$. The energy scale of the modification is denoted by $\alpha$, and the factor of $\mathcal{N}$ makes explicit the requirement that the modification's strength must scale with the size of the system. As explained above, $\hat{\chi}_E$ is an equally weighted (diagonal) distribution over all possible energy eigenstates with energies in the microcanonical interval, $[E_\psi-\sqrt{V_\psi},\,E_\psi+ \sqrt{V_\psi}]$ (set at initial times). 

While individual systems remain pure and evolve via Eq.~\eqref{Eq:thermFinal}, the master equations for the evolution of ensembles is given by:
\begin{align}\label{Eq:Therm_master_Final}&\hbar\,\frac{\partial\hat{\rho}}{\partial\,t} = 
-i\left[ \hat{H} \,,\, \hat{\rho} \right] + \alpha\,\mathcal{N}\,\,\Lambda_\chi (\,\hat{\rho}\, ),\\&\ \Lambda_\chi (\,\hat{\rho}\, )= \sum_{\mu,\nu}  \,\chi^\mu_E\, \hat{{L}}_{\mu\nu}\, \hat{\rho}\,\hat{{L}}^\dagger_{\mu\nu} - \,\hat{\rho}.\nonumber
\end{align} 
The above master equations evolve density (statistical) operators via the GKSL generator, $\Lambda_\chi$ which may be further simplified as $\Lambda_\chi (\hat{\rho}) \equiv \hat{\chi}_E - \hat{\rho}$ showing explicitly that it does not arise from a Martingale process. Together, Eq.~\eqref{Eq:thermFinal} and Eq.~\eqref{Eq:Therm_master_Final} realize a physically viable model of objective quantum thermalization (OQT) for isolated and macroscopic quantum systems. Since $\mathcal{N}$ scales with the system size, the effect of the modifications is larger for more macroscopic systems, thus admitting a quantum-to-classical transition with end states converging to classical, equilibrated configurations~\cite{aritroPhD,Bassi_03_PhyRep}. 

When the (commuting) limits, $\alpha\to0$ and $\mathcal{N}\ll\infty$ hold, standard quantum dynamics is recovered. When $\mathcal{N}\to\infty$, the limits do not commute in the dynamics of Eq~\eqref{Eq:Therm_master_Final}. In the case that $\lim_{\displaystyle \mathcal{N}\to\infty}\,\,\lim_{\displaystyle \alpha\to0}$, i.e the thermodynamic limit is taken later, standard quantum mechanics is again recovered. However, when the thermodynamic limit is approached first, i.e. $\lim_{\displaystyle\alpha\to0}\,\,\lim_{\displaystyle\mathcal{N}\to\infty}\,$, any tiny $\alpha$ can yield observable effects in the dynamics, similar to spontaneous symmetry breaking~\cite{Wezel_SSBlecturenotes,aritro3,aritro2,aritroPhD,aritro1,WezelBerry}.

To show states evolve via Eq.~\eqref{Eq:Therm_master_Final} to equilibrium, we may readily compute the (squared) Hilbert-Schmidt distance, $D_\chi = \mathrm{Tr} [ (\hat{\rho}-\hat{\chi}_E)^2  ] $. Using Eq.~\eqref{Eq:Therm_master_Final}, we find $\dot{D_\chi}=-2\,\tilde{\alpha}\,D_\chi$  (where $\tilde{\alpha}:={\alpha\,\mathcal{N}}/{\hbar}$), showing that the Hilbert-Schmidt distance exponentially decreases over time, ${D_\chi}={D_\chi}(0)\exp(-2\,\tilde{\alpha} \,t)$, where ${D_\chi}(0)$ is the distance at initial times. This shows that all states $\hat{\rho}$ which are not already the equilibrium steady states, approach $\hat{\chi}_E$. It also implies that quantum Poincar\'e recurrences or analogous entanglement reversal events~\cite{Mori_2018_ETH_rev,Stanf_Enc_phil24} do not occur for thermodynamic systems undergoing the OQT dynamics. Since microscopic correlations are also lost in the ensembles at long times, the reliability of various mean field approximations such as Boltzmann's Stosszahlansatz may be defended in the OQT dynamics in regimes  where the modifications are significant. It further implies, that the (von Neumann) entropy is non-decreasing for all states, as shown presently. 

Note that the stochastic nature of the dynamics of each ensemble member (following Eq.~\eqref{Eq:thermFinal}), itself, generates entropy for ensembles, due to the stochasticity of the individual trajectories, similar to objective collapse models~\cite{aritro2}. This is seen readily by noting the evolution of the von Neumann entropy, $S=-\mathrm{Tr}[\hat{\rho}\log\hat{\rho}]$, for ensembles evolving via Eq.~\eqref{Eq:Therm_master_Final}. The change in the entropy is given by, $\dot{S}=-\mathrm{Tr}\big[\log\hat{\rho}\frac{\partial}{\partial t}\hat{\rho}\big]$ and using Eq.~\eqref{Eq:Therm_master_Final}, we find (with $\alpha\,\mathcal{N}/\hbar =1$): 
$$
\dot{S} = \mathrm{Tr}\bigg[\big(\hat{\rho}-\hat{\chi}_E\big)\log\hat{\rho}\bigg] = \mathbb{D}[\,\hat{\chi}_E\,||\,\hat{\rho}\,] + \mathbb{H}_\chi \,- S.
$$ 
Here, $\mathbb{D}[\,\hat{\chi}\,||\,\hat{\rho}\,]= \mathrm{Tr}[\hat{\chi}\log\hat{\chi}\,]-\mathrm{Tr}[\hat{\chi}\log\hat{\rho}]$ (if $\mathrm{supp}(\hat{\chi})\subset\mathrm{supp}(\hat{\rho})$, or $+\infty$ otherwise) is the Umegaki quantum relative entropy, which is non-negative and decreases under CPTP maps~\cite{Umegaki_2,Spohn_Umegaki} such as Eq.~\eqref{Eq:Therm_master_Final}. In the above expression, $\mathbb{H}_\chi=-\mathrm{Tr}[\hat{\chi}_E\log\hat{\chi}_E]=\log \Omega$, which is the von Neumann (Shannon) entropy of the associated microcanonical distribution and $\Omega$ denotes its dimension, which is also the number of accessible micro-states at equilibrium. Since $\mathbb{H}_\chi>S$, necessarily for all non-equilibrium states, the above expression implies that we have, $\dot{S}>0$ for all non-equilibrium states and $\dot{S}\ge0$ generically. Thus the von Neumann entropy is always non-decreasing which shows a modified quantum H-theorem and the emergence of the second law of thermodynamics, independent of considerations such as the specifics of the system, its initial state, preferred observables or other agent-dependent restrictions.

Finally, by endowing dimensions to the von Neumann entropy with the Boltzmann constant, $k_B$, we obtain the analogous thermodynamic entropy for ensembles, $S_{\mathrm{Th}}=k_B\,S$. Since Eq.~\eqref{Eq:Therm_master_Final} ensures that at long times, $\hat{\rho}\rightarrow\hat{\chi}_E$, the von Neumann entropy, $S\to\mathbb{H}_\chi=\log \Omega$, showing that at equilibrium, we obtain the standard Boltzmann thermodynamic entropy, $S_{\mathrm{Th}}=k_B\,\log\Omega$, where $\Omega$ is the number of accessible micro-states in the microcanonical ensemble. This is congruent with the standard definition of micocanonical equilibrium, wherein the system may be in each admissible microstate (within the microcanonical interval) with equal probability~\cite{PATHRIA2011}. Thus, we obtain the expected microcanonical (equilibrium) statistical mechanics from the OQT model at long times, for generic thermodynamic systems. Since the total ensemble approaches an appropriate microcanonical distribution, all sub-system observables also thermalize appropriately~\cite{Goldstein_cannon_MC,PATHRIA2011}, which implies that, when relevant, using standard procedures of statistical mechanics, a canonical ensemble and a partition function may be also be constructed~\cite{PATHRIA2011}. Thus, at long times, the usual ensembles of equilibrium statistical mechanics are recovered.

Summarizing, the OQT model predicts that macroscopic quantum systems thermalize generically, at the level of their pure micro-states and such a notion of thermalization is independent of further considerations on its space of observables or states. Hence, the model in Eq.~\eqref{Eq:thermFinal} and Eq.~\eqref{Eq:Therm_master_Final} showcase an objective notion of quantum thermalization, one independent of all agent-specific arguments or approximations. The OQT model also admits a notion of irreversibility for each individual system through the dynamics of Eq.~\eqref{Eq:thermFinal}. Since each ensemble member remains pure (Eq.~\eqref{Eq:thermFinal} maps pure states to pure states), the von Neumann entropy in the individual system do not increase. For the entire ensemble, as shown above, the von Neumann entropy does increase and appropriately maximizes in each setting to the Boltzmann thermodynamic entropy (with a factor of $k_B$). 

Contrary to other approaches towards resolving the QTP, in our approach, the entropy increase in ensembles is associated to the underlying stochastic dynamics of each individual system. Further, for individual systems, there is a well-defined notion of distance from any particular equilibrium measure, whether it is $\hat{\chi}_E$ above, or any other coarser measure. Thus, this framework allows us to account for the expectation that generic thermodynamic quantum systems should thermalize objectively (at the level of ensembles), while clarifying that individual systems, at long times fluctuate near equilibrium.

Furthermore, since all observable expectation values thermalize (objectively) according to the modified dynamics of Eq.~\eqref{Eq:Therm_master_Final}, observable deviations from the predictions of standard quantum theory are expected to be realized in concrete thermodynamic systems, as alluded to before. Although a careful investigation of such deviations would rely crucially upon the concrete Hamiltonian of the situation, already in generic thermodynamic systems, the dynamics ensures that observables unrelated to any conserved quantities, which are not expected to thermalize, such as \textcolor{black}{generic} non-ETH observables will indeed thermalize at long times, while showcasing non-standard transient effects within its thermalization timescales.

This may be seen by analytically treating Eq.~\eqref{Eq:Therm_master_Final} in the interaction picture. Denote the interaction picture density operator as $\hat{\rho}_I(t)=\hat{U}^\dagger_t\,\hat{\rho}(t)\,\hat{U}_t$ where we have retained the time argument in the Schr\"odinger picture density operator, $\hat{\rho}(t)$, for clarity and $U_t:=e^{-\frac{i}{\hbar}\hat{H}t}$. Similarly other operators in the interaction picture read $\hat{O}_I(t):=\hat{U}^\dagger_t\,\hat{O}(t)\,\hat{U}_t$. In the interaction picture, Eq.~\eqref{Eq:Therm_master_Final} takes the form, $\partial_t\hat{\rho}_I(t)=\tilde{\alpha}\Big(\hat{\chi}_E-\hat{\rho}_I(t)\Big)$ with $\tilde{\alpha}={\alpha\,\mathcal{N}}/{\hbar}$ and its solution in the Schr\"odinger picture reads, $$\hat{\rho}(t)=e^{-\tilde{\alpha} t}\,\hat{U}_t \,\hat{\rho}(0)\,\hat{U}_t^\dagger \,+ \, (1-e^{-\tilde{\alpha} t})\hat{\chi}_E.$$ This again shows that any initial state of any (thermodynamic) system eventually attains the corresponding microcanonical equilibrium state, $\hat{\chi}_E$ at times $t\gg\tilde{\alpha}^{-1}$. For generic observables, $\hat{O}$, the time evolution of its expectation values, $\langle\hat{O}\rangle(t):=\mathrm{Tr}[\hat{O}\,\hat{\rho}(t)]$, is given by:$$\langle\hat{O}\rangle(t)=\langle\hat{O}\rangle_\chi+e^{-\tilde{\alpha} t}\Big(\langle\hat{O}\rangle_0\,(t)-\langle\hat{O}\rangle_\chi\Big).$$ Here $\langle\hat{O}\rangle_0\,(t) = \mathrm{Tr}[\hat{O}\,\hat{U}_t \,\hat{\rho}(0)\,\hat{U}_t^\dagger]$ is the un-modified expectation (valid for microscopic systems with $\tilde\alpha\to0$) and $\langle\hat{O}\rangle_\chi=\mathrm{Tr}[\hat{O}\,\hat{\chi}_E]$ is the thermalized expectation value \textcolor{black}{after the system achieves microcanonical equilibrium} 

\textcolor{black}{It is interesting to note that Ref.~\cite{B_new_Dabelow2020RelaxationTheory} obtains similar equations for the observable expectation values as above and uses it to study the experimental results of Ref.~\cite{A_New_Trotzky2012ProbingRelaxation}. However, crucially, the results of Ref.~\cite{B_new_Dabelow2020RelaxationTheory} are obtained by averaging over an ensemble of different systems with their Hamiltonian containing a weak random perturbation, in the framework of random matrix theory and using super-symmetric methods~\cite{Haake2018_QChaos}, apart from multiple further approximations. Some of these include the assumption that the system initially resides within a certain small interval of energies away from the ground state, the interval having approximately constant mean level spacing independent of the random perturbation, various typicality arguments which lose validity for specific systems, the restriction of observables based on the assumption that the systems follows ETH and so on (see supplementary materials of Ref.~\cite{B_new_Dabelow2020RelaxationTheory} for explicit details).}

\textcolor{black}{It is therefore striking that similar results arise readily within our OQT framework, without approximations, motivating further investigation of applications in physically relevant many-body problems showing irreversibility and randomness, which are tackled via random matrix theory. Further, this also shows that the OQT model predicts experimentally relevant deviations and may be falsified in future experiments designed towards probing non-ETH observables with controlled initial state preparations.} Indeed, since the thermalization timescales are expected to scale inversely with $\alpha\,\mathcal{N}$, the OQT dynamics of macroscopic systems deviate significantly and they thermalize much more rapidly in comparison to microscopic few body systems which remain practically indistinguishable from standard quantum theory. Thus, in the mesoscopic regime, controlled experiments could bear witness to OQT dynamics and deviations from standard quantum dynamics.

\begin{figure}[t]
\hspace{-1.1cm}
\includegraphics[width=\columnwidth]{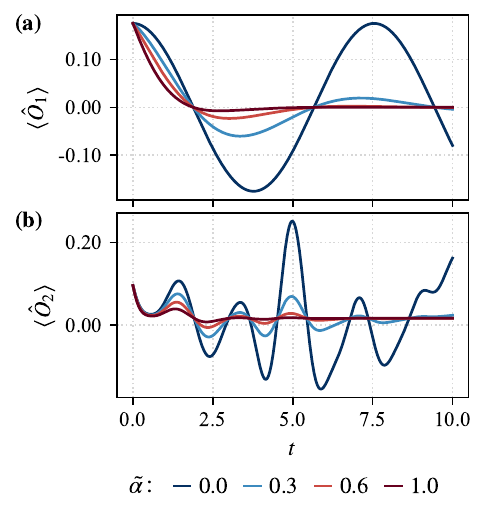}
 \vspace{-0.5cm}
\caption{Time evolution of observable expectation values, $\langle\hat{O}\rangle=\mathrm{Tr}[\hat{\rho}\,\hat{O}]$, where $\hat{\rho}$ follows Eq.~\eqref{Eq:Therm_master_Final} and the effective strength of the modification, $\tilde{\alpha}:=\alpha\,\mathcal{N}/\hbar$ is interpolated. When $\tilde{\alpha} = 0$ (dark black), standard quantum theory is recovered, while $\tilde{\alpha} >0$ (dark red, orange and light black in decreasing order) is the OQT regime which show deviations from the purely oscillatory effects of standard quantum dynamics while observable expectation values thermalize. {\fontseries{bx}\selectfont (a)}  Evolution of expectation values of $\hat{O}_1:= |i\rangle\langle j| + |j\rangle\langle i|$ where $i,j$ label energy eigenstates within the microcanonical subspace. Note, $\hat{O}_1$ thermalizes although it is not an ETH observable. {\fontseries{bx}\selectfont (b)} Evolution of expectation values of $\hat{O}_2$, a random Hermitian observable, chosen to showcase generic deviations from standard quantum predictions. Numerical analysis used $\mathrm{dim}\,\mathcal{H}=25$ (complex) with time-steps, $dt=10^{-4}$. A single generic Hamiltonian, $\hat{H}$, was chosen with random discrete spectrum in $[0.0,10.0]$, with a regular mean level spacing and standard deviation $0.01$. The initial state was a random pure state constituting a Gaussian in the energy basis amplitudes, its mean being far from the ground state and a large spread with standard deviation $0.2$. As shown analytically in the text, these results are valid quite generically.}
\label{fig1}
\end{figure}

Some of these deviations are visualized in Fig.~\ref{fig1} where we numerically investigate Eq.~\eqref{Eq:Therm_master_Final} in a $25$-dimensional (complex) Hilbert space, with a full-rank (non-trivial) \textit{single} Hamiltonian possessing random energy-levels with a regular mean spacing. This example is chosen to emulate a non-integrable quantum many-body Hamiltonian which are known to be well modelled by random matrices~\cite{ETHREview}. A random (normalized) pure initial state was sampled, with amplitudes in the energy basis being a truncated Gaussian distribution and mean energy far from its ground state energy. An appropriate target microcanonical ensemble was numerically identified using the prescription detailed before and the initial state, which was subsequently evolved numerically using Eq.~\eqref{Eq:Therm_master_Final}. The evolution of two observable expectation values are displayed, the first being $\hat{O}_1:= |i\rangle\langle j|+|j\rangle\langle i|$ (see Fig.~\ref{fig1}.\,{\fontseries{bx}\selectfont (a)}) which is not an ETH observable owing to its non-negligible off-diagonal terms ($i,j$ label energy eigenstates within the chosen microcanonical subspace). Note that observables like $\hat{O}_1$ are commonly utilized in quantum optics and spectroscopy, the analysis of transition rates in atomic and molecular systems as well as in phase coherence measurements such as in interferometry and tomography~\cite{NielsenChung11,HarocheRaimond2006}. The other observable, $\hat{O}_2$, is a full-rank (non-trivial) random Hermitian observable with maximum eigenvalue in the energy basis normalized to unity and was chosen to showcase generic deviations from standard quantum dynamics (see Fig.~\ref{fig1}.\,{\fontseries{bx}\selectfont (b)}). 

The expectation values of both observables display anomalous relaxation behavior and significant deviations from the oscillatory motifs expected generically in observable expectation values of closed systems in standard quantum theory. When the effective coupling $\tilde{\alpha}=\alpha\, \mathcal{N}/\hbar$ is zero (dark black), standard quantum dynamics is recovered whereas when $\tilde{\alpha}>0$ (shown in dark red, orange and light black in decreasing strength of $\tilde{\alpha}$), all observables thermalize according to the OQT dynamics. In other words, for systems in the thermodynamic limit when the modifications become significant, the spontaneous emergence of irreversibility ensures non-standard evolution, exposing the possibility of experimental falsification. \textcolor{black}{As shown in Fig.~\ref{fig1}, such deviations may be discernible in future experiments, particularly in the context of rapidly advancing many-body quantum simulators~\cite{A_New_Trotzky2012ProbingRelaxation,C_New_Liu2026Prethermalization,D_new_Orioli2018RelaxationIsolatedDipolar, Bernien_2017_51simul}, employing protocols such as quantum quenches~\cite{Mori_2018_ETH_rev,ETH_typ_MBL_GGE}.} In these settings, observables may thermalize even when they are not expected to do so, including non-ETH observables, unrelated to conserved quantities, with thermalization times that scale inversely with system size. Further physical input may be used towards building more finer dedicated protocols discerning such deviations in mesoscopic systems; these are left for future investigations.

Finally, notice that during objective thermalization, detailed information of the initial state is necessarily lost, so as to achieve equilibrium; whereas in any physically viable objective collapse model, the information of the initial state must necessarily remain preserved (via a Martingale condition) such that Born's rules hold at all times and super-luminal signalling is forbidden~\cite{aritroFTL,aritroPhD,Gisin:1989sx}. More precisely, in contrast to objective collapse theories, although the form of Eq.~\eqref{Eq:thermFinal} presents a stochastic Schr\"odinger equation, the crucial fact is that the analogous collapse or thermalization operators are not Hermitian and further the probability weights in the analogous (collapse or thermalization) bases are not Martingale processes~\cite{aritroPhD,aritro2,aritro3,Bassi_03_PhyRep}, which seemingly opens up the possibility of superluminal signalling due to the lack of Born's rules~\cite{aritroFTL,Gisin:1989sx,Bassi2015}. However, as will be demonstrated shortly, superluminal signalling is not allowed in the OQT model, ultimately due to the FDR and (local) linear master equations. 

Setting these causality issues aside for the moment, it is clear that the Martingale property of objective collapse models imply very different observable consequences as compared to the information scrambling dynamics of the OQT model. Concretely, the projections of the probability weights of the initial state in the collapse basis, onto the energy basis, necessarily remain preserved in objective collapse models. In contrast, in the OQT model, the same probabilities in the energy basis lose all relative ratios and equalize with support in the microcanonical sector. Thus, in general, the OQT model necessarily yields distinct qualitative and quantitative predictions as compared to objective collapse models (also see Appendix.~\ref{App.1} on the inability of collapse models in admitting standard thermodynamic equilibrium states). Crucially, both the mathematical structure and the qualitatively distinct physical implications then indicate that objective thermalization and objective collapse should be regarded as separate yet coexisting and competing mechanisms, suggesting a balance between classical ordering, via objective collapse and disordering through objective thermalization, as discussed in Sec.~\ref{Sec4:SUI}.

Another subtle difference is that the OQT model requires a specification of an average energy interval, denoted $E_\psi$-sector, functioning as the microcanonical ensemble, which in turn controls the structural specifications of Eq.~\eqref{Eq:thermFinal} and Eq.~\eqref{Eq:Therm_master_Final}. Once the $E_\psi$-sector is specified, Eq.~\eqref{Eq:thermFinal} results in a scrambling dynamics for each ensemble member (controlled by the instances of the stochastic contributions), while the noise-averaged ensemble distribution approaches $\hat{\chi}_E$. Thus, the thermalization model is not apriori defined for all setups, but is obtained uniquely in each case via an analysis of the Hamiltonian, similar to the construction of recently proposed objective collapse models rooted in spontaneous symmetry breaking~\cite{aritro1,aritro2,aritro3,aritroPhD}.

Note that an important shortcoming of the OQT model is that it is driven by uncorrelated white noise, which cannot constitute a physical process. However, the model may be viewed as an effective Markovian description of a colored noise driven model of OQT. This mapping is made precise via the so called multiscale noise homogenization procedure---a temporal renormalization scheme on the space of colored noise driven models~\cite{aritro2,aritro3,aritroPhD,aritro_Ito_Strat}. The key take away of such a procedure would be that multiple colored noise models would ultimately flow to Eq.~\eqref{Eq:thermFinal} and Eq.~\eqref{Eq:Therm_master_Final} ~\cite{aritro2,aritro3, aritroPhD,aritro_Ito_Strat}. Further, the model is constructed for countable spectrum of energy eigenstates, and may be generalized to the continuum setting by employing multi-parameter Wiener processes or space-time white noise~\cite{aritro3,aritroPhD}. The analysis of physical consistency in such continuum and colored noise driven models is left for future studies. 

Further, analogous to open problems in relativistic collapse models~\cite{Bassi_rela_Jones_2021}, the construction of OQT models in the relativistic regime remains another challenge and is left for future investigations. In this context, the formulation in Eq.~\eqref{Eq:thermFinal} via the $E_\psi$-sectors, may be particularly advantageous towards treating quantum systems in the strict thermodynamic limit, as well as quantum field theories, where these $E_\psi$-sectors map on to disjoint super-selection sectors, each sector furnishing a reference state and its own representation of the observable algebra~\cite{Landsman2017}.

\subsection{\textbf{Superluminal signalling}}\label{Sec3B}Having constructed a physically viable OQT model, we will now show that it does not allow faster than light communication or signalling. Unphysical scenarios allowing superluminal signalling is a common issue encountered in modified quantum theories~\cite{Gisin:1989sx,Bassi2015,aritroFTL}. It is linked to the possibility that spatially separated parties can infer each others actions faster than light, given the modified dynamics on the entire Hilbert space. Residual non-linearites in the master equation are one key indication of such possibilities, since ensembles evolving via non-linear master equations do not remain equivalent and allow superluminal signalling~\cite{Gisin:1989sx,Bassi2015,aritroFTL}. Since the master equation in Eq.~\eqref{Eq:Therm_master_Final} is linear and of the GKSL form, it is assured that there is no superluminal signalling as long as the thermalization operators, $\hat{L}_{\mu\nu}$ have local support~\cite{aritroFTL}. We now re-verify this via the necessary and sufficient condition and witness of superluminal signalling in Ref.~\cite{aritroFTL}. 

Consider the usual setup of Alice and Bob in their spacelike separated labs with a tensor product Hilbert space, $\mathcal{H}_A\otimes\mathcal{H}_B$. They have  access to an ensemble of entangled particles. One may imagine that Alice additionally possesses a macroscopic quantum system (like a quantum gas in a box) and Alice's share of the entangled particle impinges on it. While Alice's system equilibrates using the modified dynamics of Eq.~\eqref{Eq:thermFinal}, Bob may perform local operations and measurements on his entangled particle. The spacelike separation implies that the Hamiltonian describing the evolution of the joint state possesses no interactions between the setups of Alice and Bob. For an ensemble of such setups, if Bob notices any change in his measurements due to Alice's share undergoing non-trivial dynamics, Alice has effectively sent a signal to Bob faster than light. We will now see how this is disallowed given that the thermalization of Alice's system is local. 

Let Alice and Bob share an ensemble of arbitrary entangled states of the form, $\ket{\Psi_{AB}}:=\sum_{\mu,\sigma} \psi_{\mu,\sigma}\,\ket{\mu_A}\otimes\ket{{\sigma_B}}$. While Bob has access to states, indexed by $\sigma$, ($\,\ket{\sigma_B}\in\mathcal{H}_B\equiv\mathbb{C}^M$ ) which he can projectively measure, Alice has access to an $N$-state macroscopic system ($\,\ket{\mu_A}\in\mathcal{H}_A\equiv\mathbb{C}^N$) entangled with Bob's share. Let Alice's macroscopic state now thermalize \textit{locally} following the dynamics of Eq.~\eqref{Eq:thermFinal} and Eq.~\eqref{Eq:Therm_master_Final}. This implies that in particular, $\hat{G}_{\chi}\equiv\hat{G}^A_{\chi}\otimes\hat{\mathbb{I}}_B$, and the operators, $\hat{L}_{\mu\nu}\equiv\hat{L}^A_{\mu\nu}\otimes\hat{\mathbb{I}}_B$, where $\ket{\mu_A},\ket{\nu_A}\in\mathcal{H}_A$ are the energy eigenstates of Alice's local setup. 

Since Bob has access to local projective measurements and operators ($\hat{\mathbb{I}}_A\otimes \hat{O}_B$), he may only access the reduced state $\hat{\rho}_B=\mathrm{Tr}_A[\hat{\rho}_{AB}]$~\cite{aritroFTL}. Here, $\hat{\rho}_B$ is the reduced density matrix for Bob which is obtained after a partial trace ($Tr_A[.]$) procedure on the entire density matrix $\hat{\rho}_{AB}$. The change in Bob's ensemble averages (for local observables) due to Alice's system thermalizing is then given by $\frac{\partial\hat{\rho}_B}{\partial\,t} = Tr_A\big[\frac{\partial}{\partial\,t}\hat{\rho}_{AB}\big]=0
$ unconditionally, where $\hat{\rho}_{AB}$ evolves via Eq.~\eqref{Eq:Therm_master_Final} with local thermalization operators as described above. This clearly shows that a spatially extended macroscopic entangled state may objectively thermalize locally, however superluminal signalling is prohibited within the dynamics of Eq.~\eqref{Eq:thermFinal} and Eq.~\eqref{Eq:Therm_master_Final}. 

If $\frac{\partial\hat{\rho}_B}{\partial\,t} \neq 0$, typically due to non-linear expectation values in the master equation, then indeed Bob would have been able to distinguish his ensemble before and during objective thermalization, allowing Alice to effectively signal faster than light. Clearly this is not possible within the proposed model with local thermalization operators, since all non-linearities cancel given the FDR, which thus is also seen to guarantee no superluminal signalling in the theory. With these constraints satisfied, we now focus on incorporating objective collapse in the OQT model.

\section{Spontaneous Universal Irreversibility}\label{Sec4:SUI}
The objective quantum thermalization model (Eq.~\eqref{Eq:thermFinal} and Eq.~\eqref{Eq:Therm_master_Final}) and the various objective collapse models~\cite{Bassi_03_PhyRep,Bassi2013Review,aritro1,aritro2,aritro3,Wezel10} showcase the possibility of a previously unrecognized, universal \textcolor{black}{theory of stochasticity and irreversibility for quantum systems approaching the thermodynamic limit}~\cite{aritroPhD}. In particular such models modify the deterministic and time-reversal symmetric evolution of quantum systems and allow them to undergo fundamentally stochastic and irreversible dynamics, expected phenomenologically, but disallowed in standard quantum dynamics. These modifications are extensive and their strength depends on the system size, fundamentally affecting microscopic and macroscopic quantum matter differently and are thus experimentally verifiable~\cite{Bassi2013Review,Wezel_2012,Carlesso_2022}. Together, a hybrid dynamics of both objective collapse and objective thermalization is seen to resolve both the quantum measurement problem and the QTP within the same theory~\cite{aritroPhD}. 
 
In this view, the standard practice of  substituting `by hand', equilibrium states, symmetry broken states or projected states (during measurements) may be seen as a minimal, economical way of mapping onto the observed physics and not that such processes are instantaneous~\cite{aritroPhD}. Indeed, since these are inherently dynamical and non-equilibrium processes, the next refinement towards describing their physics is in the Markovian approximation. \textcolor{black}{ Barring instantaneous jumps and adhering to the various constraints, dynamics of the form described here (Eq.~\eqref{Eq:thermFinal} and Eq.~\eqref{Eq:Therm_master_Final}) constitute the only unique continuous possibility (upto unitary transformations) which result in microcanonical equilibrium in closed systems. Further, continuous models discussed in the context of objective collapse are the only unique possibilities upto a choice of collapse operators and unitary transformations~\cite{Bassi_unique,aritro3,Bassi_03_PhyRep,Bassi2013Review,aritroPhD}}. 

Following this line of thought, we now construct a unique (upto unitaries and choice of collapse operator) continuous class of models showcasing fundamental quantum stochasticity and irreversibility for systems approaching the \textit{thermodynamic limit}~\cite{Wezel_SSBlecturenotes,PATHRIA2011,Landsman2017} or equivalently, systems beyond the so called \textit{Heisenberg's cut}~\cite{aritroPhD,Landsman2017,auletta2000foundations}, or indeed, systems in the \textit{quantum-to-classical crossover regime}~\cite{aritroPhD,Bassi_03_PhyRep,Bassi2013Review}. Such terminology, although previously used in different situations, may now be seen to indicate the same regime where quantum theory requires modifications so as to account for the spontaneous emergence of classicality. In particular, such hybrid models, with generators admitting both objective collapse and objective thermalization are henceforth termed models of Spontaneous Universal Irreversibility (SUI). We now briefly describe their construction. 

SUI models consists of two distinct modifications to standard quantum dynamics, one allowing dynamical quantum state reduction or objective collapse and one allowing objective thermalization, both given by a (norm-preserving) quantum stochastic process on the entire Hilbert space $\mathcal{H}$ of the isolated system, given by:
\begin{align}
d&\ket{\psi} =\,-\frac{i}{\hbar}\hat{H}\ket{\psi}dt + d\hat{G}_R\ket{\psi}+d\hat{G}_\chi\ket{\psi}.  \label{Eq:SUI}
\end{align}
Here, as before, $\ket{\psi}$ denotes the state of a single isolated (thermodynamic) system with its standard Hamiltonian $\hat{H}$. Further, $d\hat{G}_\chi:=\sum_{\mu\nu} d\hat{G}_\chi^{\mu\nu}$, given in Eq.~\eqref{Eq:thermFinal} generates objective quantum thermalization ensuring that isolated systems approach an appropriate notion of microcanonical equilibrium, $\hat{\chi}$, given conserved quantities.

$d\hat{G}_{R}$ is the modified generator of wavefunction reduction. Various generators of different objective collapse models~\cite{Bassi_03_PhyRep,Bassi2013Review, aritro2, aritro3} may be considered, but note that all such physically viable generators are stringently constrained, requiring that both stochastic and deterministic contributions are present and related by a fluctuation dissipation relationship~\cite{aritro2,aritro3,aritroPhD}, as also seen in the OQT model before. Further, just like the OQT model (Eq.~\eqref{Eq:thermFinal} and Eq.~\eqref{Eq:Therm_master_Final}) which allows the equilibrium microcanonical distribution $\hat{\chi}$ to emerge in ensembles at long times, a diagonal distribution, $\hat{\rho}_\mathrm{B}$,  corresponding to Born's rules in a \textit{particular basis}, emerges due to the role of $d\hat{G}_{R}$ at long times (in Eq.~\eqref{Eq:SUI} with $\hat{H}=0$ and $d\hat{G}_{\chi}=0$). 

The various objective collapse models impose this preferred basis of collapse, determining the form of $d\hat{G}_{R}$. The well known Continuous Spontaneous Localization (CSL)~\cite{Ghirardi_90_PRA,Bassi_03_PhyRep} and the Diosi-Penrose (DP) models~\cite{Diosi_87_PLA} specifies the eigenstates of the mass density operator as its preferred basis, while the (non-continuous) Ghirardi–Rimini–Weber (GRW) model~\cite{Ghirardi_1986} and quantum mechanics with universal position localization (QMUPL)~\cite{diosi_1989} utilizes the position basis. Note that, all these models specify a preferred basis, chosen apriori by their proposed mechanisms and their dynamics differ and yield differing predictions accordingly. However, the competition between $d\hat{G}_{R}$ and $d\hat{G}_{\chi}$ will ensure steady state distributions with both thermalization and quantum state reduction, as seen from the generic form of the master equations described shortly.

We now focus on a recently established objective collapse model termed Spontaneous Unitarity Violations (SUV) ~\cite{aritro1,aritro2,aritro3,WezelBerry,Wezel10,Wezel_2009,Wezel_2008,aritroPhD} which differs from the above collapse models in various crucial respects. SUV is motivated by extending spontaneous symmetry breaking~\cite{Wezel_SSBlecturenotes,Brauner2024Effective_SSB} to the non-equilibrium regime and its dynamics allow macroscopic quantum objects to spontaneously break their symmetries, disallowed in standard quantum dynamics~\cite{WezelBerry,Wezel10}. \textcolor{black}{Note, in a closed setup evolving unitarily, spontaneous symmetry breaking generated by symmetry breaking perturbations must be fine tuned to yield Born's rules and further, can atmost lead to oscillations with no steady symmetry broken state being reached. Further in an open setting, symmetry breaking always imply macroscopic superpositions wherein the system undergoing symmetry breaking and the environment supplying the perturbations remain entangled. Arguments based on decoherence and tracing out the environment thus cannot explain symmetry breaking in single systems. Indeed, both cases result in the persistence of macroscopic superpositions which lead back to the quantum measurement problem~\cite{aritroPhD,Wezel10,Wezel_SSBlecturenotes,WezelBerry,aritro1,aritro2,aritro3}. The SUV model resolves this via non-unitary symmetry breaking perturbations, allowing systems to dynamically collapse onto macroscopically different, symmetry broken sectors while adhering to Born's rules.} The same model applied to a measurement setting---composed of a system being measured, macroscopic measurement devices and an environment---allows the entire setup to undergo quantum state reduction to a classical state, wherein the device breaks a symmetry, allowing measurements to be performed~\cite{aritro1,aritro2,aritro3,aritroPhD}, such as a pointer breaking translation symmetry. Thus within the SUV mechanism, quantum state reduction and spontaneous symmetry breaking are two descriptions of the same irreversible and inherently random phenomenon for systems approaching the thermodynamic limit~\cite{aritro3,aritro2,WezelBerry,Wezel10,Wezel_2009,Wezel_2008}. 

Unlike other collapse models~\cite{Bassi_03_PhyRep,Bassi2013Review}, there is no preferred basis chosen apriori, instead, in each situation the SUV model uniquely determines the symmetry broken basis of the setup's Hamiltonian as its basis of wavefunction reduction ~\cite{aritro2,aritro3,Wezel10,Wezel_2009,Wezel_2008}. This ensures that expectation values of order parameters (macroscopic observables capable of distinguishing symmetry broken configurations ~\cite{Wezel_SSBlecturenotes}) achieve a determinate value in each dynamical realization. This mechanism thus allows quantum state reduction of a superposition of macro-states to a single (classical) macro-state, without affecting the relative probabilities between the micro-states (in the realized macrostate). Further, in the Markovian regime, the SUV model reduces to an effective stochastic Schr\"odinger equation, collapsing in the symmetry broken basis~\cite{aritro3}. The discrete (Markovian) SUV generator is of the form:
\begin{align}
\,d\hat{G}_R&\ket{\psi} =\,- \frac{\mathcal{J}\mathcal{N}'}{2\hbar}\, \sum_k  \left(\hat{\mathbb{P}}_k - \langle\hat{\mathbb{P}}_k\rangle  \right)^2\ket{\psi} dt \notag \\
 &+  \,\sqrt{\frac{\mathcal{J} \mathcal{N}'}{\hbar}} \sum_k \left(\hat{\mathbb{P}}_k - \langle\hat{\mathbb{P}}_k\rangle  \right)  \ket{\psi} dW^k_t. \nonumber
\end{align}
Here $\mathcal{J}$ determines the strength of the collapse dynamics, $\mathcal{N}'$ increases with the system size. The projection operators $\hat{\mathbb{P}}_k=\ket{k}\bra{k}$ projects onto macroscopically distinguishable sectors of the Hilbert space, such as symmetry broken configurations, determined from $\hat{H}$, which crucially allows the SUV approach to model quantum state reduction and spontaneous symmetry breaking as two descriptions of the same irreversible and inherently random phenomenon~\cite{aritro3,aritro2,WezelBerry,Wezel10,Wezel_2009,Wezel_2008}. 
Since all three stochastic and irreversible phenomena---spontaneous symmetry breaking, quantum measurements and thermalization---occur only for systems approaching the thermodynamic limit~\cite{Wezel_SSBlecturenotes,Bassi_03_PhyRep}, a SUI model constructed using the SUV model~\cite{aritro2,aritro3,aritro1,aritroPhD} and the OQT model (Eq.~\eqref{Eq:thermFinal} and Eq.~\eqref{Eq:Therm_master_Final}) \textcolor{black}{can function together seamlessly and yields a universal theory describing the quantum-classical crossover physics in the non-relativistic regime. }

This is seen clearly by noting that the SUV collapse operators, $\hat{\mathbb{P}}_k=\ket{k}\bra{k}$ project onto macroscopically distinguishable configurations, without affecting the dynamics of the internal (relative) degrees of freedom i.e. the micro-states~\cite{aritroPhD,aritro3,aritro2,Wezel10}. \textcolor{black}{For example, in a crystal, the SUV (continuum) projection operators project onto a subspace with all micro-states with a definite position of the center of mass, which functions as the order parameter, resulting in just the center of mass localizing without affecting the internal degrees of freedom (relative coordinates)~\cite{aritroPhD,aritro3}.} On the other hand, $d\hat{G}_\chi$, as defined before results in micro-canonical equilibrium, while keeping conserved quantities conserved, which also include order parameters such as the total momentum, the total magnetization, total angular momentum and so on. Thus while the SUV dynamics will localize these (global) macroscopic order parameters fixing a classical macro-state, the OQT dynamics effectively ensures that the system traverse all the micro-states present within each sector which correspond to macroscopically distinguishable configurations. Thus, at long times, the hybrid dynamics result in an ensemble of closed quantum systems approaching both localized (classical) macrostates (for each ensemble member) while adhering to microcanonical equilibrium. 

Such a dynamics is distinct from the competition encountered in SUI models constructed with other collapse mechanisms where the collapse basis is fixed, although the form of the equations remain the same with the $\hat{\mathbb{P}}_k$ operators being replaced by the corresponding collapse operators of the model considered~\cite{Bassi_03_PhyRep,Bassi2013Review}. In all such SUI models however, a single isolated quantum system in the thermodynamic limit, will dynamically converge to classical, equilibrated configurations. The steady states in these models may be different and could be used to experimentally constrain them further. 

The equilibrium steady states are obtained from the corresponding master equations for ensembles of isolated systems evolving via Eq~\eqref{Eq:SUI}, which is a (linear) GKSL master equation with two contributions:
\begin{align}
    &\hbar \frac{\partial \hat{\rho}}{\partial t} = -i \left[ \hat{H} \,,\, \hat{\rho} \right] + \mathcal{J}\,\mathcal{N}'\,\,\Lambda_R (\,\hat{\rho}\, )+ \alpha\,\mathcal{N}\,\,\Lambda_\chi (\,\hat{\rho}\, ).
    \label{Eq:masterSUI}
\end{align} 
Here, $\Lambda_R (\,\hat{\rho}\, ):=  \sum_k \hat{\mathbb{P}}_k \hat{\rho} \hat{\mathbb{P}}_k - \hat{\rho}$ is the SUV generator of wave-function reduction in a symmetry broken basis specified by the set $\{\hat{\mathbb{P}}_k\}_{k\in\mathbb{N}}$, which allows an ensemble averaged description of both spontaneous symmetry breaking and quantum measurements in their appropriate settings~\cite{aritroPhD,aritro3,aritro2,aritro1,Wezel10,Wezel_2009,WezelBerry}. As explained, $\Lambda_R$ generates ordering at the level of macroscopic order parameter observables, without affecting the ratio of probability amplitudes between micro-states corresponding to the same classically ordered macro-state. It leads to increase in von Neumann entropy and order parameter expectation values acquire determinate values corresponding to classical configurations, for example, there may spontaneous magnetization in a ferromagnet or spontaneous localization of the center of mass of a pointer~\cite{aritroPhD,aritro3,aritro2,aritro1,Wezel10,Wezel_2009,WezelBerry}. In the thermodynamic limit, any change in the average energy ($E=\mathrm{Tr}[\hat{\rho}\hat{H}]$) is highly suppressed by the SUV generator, which may be seen by noting that for time independent \textcolor{black}{Hamiltonians with local interactions}, Eq.~\eqref{Eq:masterSUI} yields $\dot{E}\propto\mathrm{Tr}[\hat{\rho}\,\Lambda_R (\hat{H} )]$. Since the $\hat{\mathbb{P}}_k$ represent projections onto macroscopically different sectors with $\sum_k\hat{\mathbb{P}}_k=\hat{\mathbb{I}}$, one finds that $\Lambda_R (\hat{H} )=\sum_{k\neq l}\hat{\mathbb{P}}_k\hat{H}\hat{\mathbb{P}}_l\to 0$ in thermodynamic systems~\cite{aritro3}. 

In Eq.~\eqref{Eq:masterSUI}, $\Lambda_\chi$ generates objective thermalization in ensembles, keeping conserved quantities preserved, as discussed before (see Eq.~\eqref{Eq:Therm_master_Final}). $\Lambda_\chi$ generates a disordering dynamics such that the system remains in the same macro-state, while effectively equalizing the probability of all accessible micro-states. As shown in Sec.~\ref{Sec3a}, it increases the von Neumann entropy for ensembles, which (upto a factor of $k_B$) converges to the Boltzmann thermodynamic entropy at equilibrium. Further, both the SUV and OQT models posses norm preservation and no-superluminal signalling, \textcolor{black}{while any change in the average energy is suppressed in the thermodynamic limit}~\cite{aritroPhD,aritro2,aritro3}, strengthening their viability as physically relevant models for the dynamics of isolated systems. These properties are ultimately fixed by the FDR relating the deterministic and stochastic contributions of the modifications and possibly hinting at their underlying origin~\cite{aritro2,aritro3,aritroPhD}.

Note that generators of other objective collapse models will also generically yield a master equation of the form in Eq.~\eqref{Eq:masterSUI}, with differing collapse operators. Together, the combined effect of $\Lambda_R$ and $\Lambda_\chi$ describe competing, thermalization-collapse dynamics with long-time steady, equilibrium distributions denoted by $\hat{\rho}_\infty$. They are obtained by the steady state condition in each setting and is given by $i \left[ \hat{H} \,,\, \hat{\rho}_\infty \right] = \mathcal{J}\,\mathcal{N}'\,\,\Lambda_R (\,\hat{\rho}_\infty)+ \alpha\,\mathcal{N}\,\,\Lambda_\chi (\,\hat{\rho}_\infty )$. \textcolor{black}{While using the SUV model as the collapse mechanism,} the $\hat{\rho}_\infty$ integrates the diagonal Born distribution, $\hat{\rho}_\mathrm{B}$, and the microcanonical distribution, $\hat{\chi}$,  allowing an unified description of quantum state reduction, spontaneous symmetry breaking and quantum thermalization, for isolated macroscopic quantum systems. Since the microcanonical distribution was identified by Maxwell, Boltzmann and Gibbs in their respective works~\cite{Sklar1993Physics,Stanf_Enc_phil24}, one may term the $\hat{\rho}_\infty$ as a hybrid Born-Maxwell-Boltzmann-Gibbs-microcanonical distribution. Such hybrid steady state distributions are also be obtained while utilizing generators of other objective collapse models. Further analysis of these steady states towards understanding quantum-to classical transitions in particular settings is left for future investigations. 

Thus, SUI models can consistently and un-ambiguously describe the irreversible, quantum-to-classical crossover dynamics, in both individual systems and ensembles. \textcolor{black}{Indeed, the irreversible dynamics in SUI models reduce quantum states of  macroscopic systems to classical localized and equilibrated states. Thus macroscopic entanglement and correlations are suppressed which may support the dynamical emergence of physics underlying widely successful mean field approximations such as the Stosszahlansatz, a premise left for future investigations. } For microscopic systems the SUI models are practically indistinguishable from standard quantum theory, but due to their differing dynamics, yields falsifiable predictions for quantum systems approaching the thermodynamic limit~\cite{Bassi2013Review,Wezel_2012,Carlesso_2022}. Further, although our considerations in this article were restricted to non-relativistic systems, ultimately, our program aims to provide a unified description of irreversibility and stochasticity in general space-times for thermodynamic quantum systems of particles and fields at the quantum-to-classical crossover regime. The construction of relativistic SUI models and the analysis of dedicated experimental protocols falsifying them are left for future studies.
\section{Conclusions}\label{Sec5:Concl}
The deterministic and time-reversal symmetric nature of standard quantum dynamics poses a foundational challenge in reconciling it with the irreversible approach of thermodynamic systems towards equilibrium, held as an axiom in equilibrium statistical mechanics. Standard resolutions employ epistemic restrictions, based on what may be practically known by specific agents. However, in such approaches, an agent-independent (objective) notion of equilibrium in isolated systems is not possible, nor is it understood how a single, isolated, macroscopic system irreversibly approaches equilibrium. Our work resolves this tension by instead advocating that quantum mechanics is an effective theory, requiring corrections to accurately describe the dynamics of thermodynamic systems. 

We proposed a minimal stochastic modification of quantum dynamics, such that for microscopic systems, the modified dynamics is practically indistinguishable from Schr\"odinger's evolution, while macroscopic systems can approach equilibrium dynamically. Concretely, the proposed objective quantum thermalization (OQT) model ensures that isolated macroscopic systems, governed by corrections scaling with its system size, evolve irreversibly toward thermal equilibrium described by a microcanonical distribution. Crucially, this approach to equilibrium remains a well defined notion for both single systems and ensembles, as well as for all possible observables or initial states of a given (macroscopic) system, transcending subjective or agent-dependent restrictions inherent in all other frameworks~\cite{aritroPhD,Jaynes_1957,Jaynes_1957_2}.

We showed that a fluctuation-dissipation relation guarantees that the OQT model is norm-preserving and avoids superluminal signalling, ensuring physical consistency. We showed that constraints ensuring energy conservation and equilibrium steady states may be simultaneously satisfied and constructed an OQT model with increasing von Neumann entropy in ensembles of isolated systems. The ensemble was shown to approach microcanonical equilibrium at long times, for thermodynamic systems, independent of further considerations. A modified H-theorem was shown to hold and the von Neumann entropy (scaled with $k_{\mathrm{B}}$) was shown to converge to the Boltzmann thermodynamic entropy at equilibrium. Further, owing to the differing dynamics in macroscopic quantum systems, the OQT model presented in this article was shown to be falsifiable via dedicated protocols in the mesoscopic regime, similar to objective collapse theories~\cite{Bassi2013Review,Wezel_2012,Carlesso_2022,Romeral_2024}. We further argued that OQT and objective collapse constitute fundamentally distinct mechanisms, leading to both qualitative and quantitative differences in their physical predictions.

However, both these mechanisms were argued to be necessary in order to recover classicality in thermodynamic limit quantum systems. We thus considered the integration of OQT with objective collapse theories, which allow a unified framework, termed spontaneous universal irreversibility (SUI); addressing both the quantum measurement problem and the quantum thermalization problem. We gave a general prescription towards their construction and focused on developing a particular SUI model which describes thermodynamic systems undergoing stochastic dynamics that drive spontaneous symmetry breaking, wavefunction reduction and objective thermalization, while remaining indistinguishable from standard quantum theory for microscopic systems. 

This hybrid model reconciles the emergence of classicality and equilibrium thermodynamics from its quantum mechanical constituents, resolving both the quantum measurement problem and the quantum thermalization problem, within the same theory. The SUI model provides a self-consistent theory of quantum irreversibility and quantum-to-classical crossover dynamics, where the interplay of stochastic collapse and thermalization dynamics results in emergent, long-time steady states, corresponding to a hybrid Born-Maxwell-Boltzmann-Gibbs-microcanonical distribution, i.e. it integrates both Born's rules and microcanonical equilibrium statistics.

The further investigation of SUI models may have deep foundational significance as well as a myriad of practical applications. \textcolor{black}{In the non-relativistic regime, the SUI models open up possibilities of finer analyses of discrepancies in thermalization timescales and characterization of noise in quantum gasses and condensed matter systems~\cite{Mori_2018_ETH_rev,aritroPhD} as well as in quantum devices such as quantum many-body simulators and quantum computers~\cite{A_New_Trotzky2012ProbingRelaxation,C_New_Liu2026Prethermalization,D_new_Orioli2018RelaxationIsolatedDipolar, Bernien_2017_51simul,aritroPhD, Preskill2018quantumcomputingin}.} 

Spontaneous symmetry breaking~\cite{Wezel_SSBlecturenotes,Brauner2024Effective_SSB} is a corner stone of modern physics, from condensed matter to the standard model of particle physics and since the SUI models provides its natural extension in the non-equilibrium regime~\cite{aritro2,aritro3,WezelBerry,Wezel_2008,Wezel_2009,Wezel10}, they are of considerable interest towards treating such quantum-to-classical transitions dynamically both in non-relativistic systems as well as in the relativistic, high energy setting, such as probed in recent collider tests of quantum foundations~\cite{High_Bell_1,Timpson23_high_bell2}. Indeed, relativistic SUI models could be used to investigate known conundrums of thermalization time scales in relativistic heavy-ion-collision physics~\cite{hypertriton} and in symmetry breaking phase transitions and non-equilibrium physics in early universe~\cite{CMB1,CMB2}. The construction of such SUI models in the relativistic regime remain an open problem with both conceptual and technical subtleties~\cite{Bassi_rela_Jones_2021}. 

Crucially, note that although we have constructed equations of motion admitting long-time steady states, integrating microcanonical equilibrium and Born's rule, our analysis, however, does not specify \textit{why} these should be the preferred steady state distributions in the first place. Nor does it expose constraints controlling the mutual competition and co-existence of both thermalization and collapse mechanisms. Thus, a key open problem is also to understand the underlying source of such stochastic modifications, which would presumably inform these concerns. In the context of objective collapse models, its origins have been previously argued to emerge due to an instability of superposed space-times of massive gravitating quantum objects~\cite{Penrose_96,penrose14}. \textcolor{black}{It has been argued that the black-hole information paradox motivates non-deterministic and non-unitary (irreversible) dynamics that map pure states to mixed states~\cite{BH_Info_col1,BH_Page1,BH_Hawking1982Unpredictability}. It has also been argued that incorporating classical gravity into a hybrid classical-quantum framework requires stochasticity and irreversibility for consistency~\cite{Oppenheim_Post_QG}.} Indeed, whether such mechanisms could realize SUI like modifications, remains an open question.

In closing, our results challenge conventional interpretations of quantum theory and the emergence of classicality and equilibrium. By positing stochasticity in the fundamental dynamics of physical systems, our framework lays bare an objective understanding of quantum-to-classical transitions and the emergence of equilibrium in both single systems and ensembles, offering a unified dynamical framework towards  describing spontaneous symmetry breaking, quantum state reduction and thermal equilibration. Our findings thus motivate and encourage a re-assessment of the foundations of physics via theoretical and experimental exploration of potential universal mechanisms which could underlie the irreversibility and stochasticity in the observed universe.
\subsection*{\emph{\textbf{Acknowledgments}}}I gratefully acknowledge discussions with J. van Wezel, T. M. Nieuwenhuizen, L. P. Hughston, N. P. Landsman, K. Hornberger,  T. P. Singh, J. Dupont,  O. Shor, F. Holik, L. Di\'osi, A. Bassi, H. Maassen, P. Grangier, M. Ozawa, U. Bhardwaj, J. Veenstra, S. Mukhopadhyay and my family, D. Mukherjee, B. Mukherjee and A. Mukherjee. I respectfully acknowledge the late Prof. K. R. Parthasarathy, whose profound contributions in quantum stochastic calculus and extensive corpus of research served as a significant source of inspiration. I am also deeply grateful to J\~n$\bar{\mathrm{a}}$n\=ananda Seva Sangha and S. M. Param$\bar{\mathrm{e}}$shwari Devi for their gracious support and valuable insights. I acknowledge support by the Open Access Publication Fund of the University of Duisburg-Essen.

\appendix
\section{Objective Collapse and Objective Thermalization: Similarities and Differences~\label{App.1}}
This appendix discusses parallels and distinctions between objective collapse models and the proposed OQT model (Eq.~\eqref{Eq:thermFinal} and Eq.~\eqref{Eq:Therm_master_Final} in Sec.~\ref{Sec3a}). We define objective collapse mechanisms as models whose steady states are diagonal density matrices $\hat{\rho}_B$ consistent with Born’s rules, with their diagonal elements preserved in the collapse basis~\cite{Bassi_03_PhyRep} (see Sec.~\ref{Sec4:SUI}). In contrast, the OQT model results in a thermal distribution $\hat{\chi}$ where all information of the initial condition is lost, except the average energy and other conserved quantities. This already indicates that collapse models and the OQT model are distinct, with both qualitative and quantitative differences.

In Refs.~\cite{DavidAlbert94,DavidAlbert2000TimeAndChance}, it has been claimed that since objective collapse mechanisms incorporate irreversibility and an arrow of time into the dynamics, they are sufficient in allowing macroscopic systems to attain thermal equilibrium without certain epistemic considerations. Further, Refs.~\cite{DavidAlbert94,DavidAlbert2000TimeAndChance} proposes that the GRW (Ghirardi-Rimini-Weber) model~\cite{Ghirardi_1986} with discontinuous jump processes can lead to thermal equilibrium, however no explicit proofs or concrete analytically worked out examples are provided. 

We now argue why such a mechanism will not result in objective thermalization in closed systems, as is the case for the OQT model (Eq.~\eqref{Eq:thermFinal} and Eq.~\eqref{Eq:Therm_master_Final} in Sec.~\ref{Sec3a}). The mechanism proposed in Refs.~\cite{DavidAlbert94,DavidAlbert2000TimeAndChance} begins with the suggestion that so-called abnormal or anti-thermodynamic states, which under standard unitary time evolution would result in apparent decrease in entropy (as seen via a restricted set of observables), are a set of measure zero within the entire state space. The GRW mechanism suppresses these abnormal states and only normal states, which under unitary time evolution result in apparent increase in entropy survive and thus eventually, thermal equilibrium is achieved. Apart from the restriction on observables, this proposal puts restrictions on the initial state of the system, as well as a restriction on viable systems--those that are thermodynamically normal. It is conceivable that systems may be prepared in such an abnormal state, close to it or a superposition of them (such an initial state may be thermodynamically distinct from normal state) and then it would never achieve thermal equilibrium within this proposal. That certain configurations constitute a measure zero set, does not imply that they may be ignored physically~\cite{Sklar1993Physics,Stanf_Enc_phil24}. Indeed, it is well known that in complex classical systems with mixed phase spaces composed of chaotic regimes with embedded small islands of regularity, the very existence of these islands can significantly change the system dynamics and its relaxation properties~\cite{Sticky_Hamiltonian,sticky_3,Sticky_ZASLAVSKY2002461}. Finally special, low-entropy initial conditions are employed~\cite{DavidAlbert2000TimeAndChance}.

Note that the role of the environment is unclear and the proposal does not constitute a complete description of closed system thermalization; it implies a restriction on the set of observables or states (see Sec.~\ref{Sec2:Lit_Rev}). Further it is not clear why collapse in the position basis would result in the emergence of thermalized ensembles in the energy basis. Crucially, it is now well established that the original GRW mechanism fails to properly accommodate identical particles and leads to energy violations~\cite{Bassi_03_PhyRep}, which further undermines the proposal. In contrast, the OQT model (Eq.~\eqref{Eq:thermFinal} and Eq.~\eqref{Eq:Therm_master_Final} in Sec.~\ref{Sec3:OQT}) conserves the energy and does not require any of the above restrictions on systems, initial states or observables.

For the sake of completeness, in this appendix, we explicitly prove the converse of the claim in Refs.~\cite{DavidAlbert94,DavidAlbert2000TimeAndChance}, in the more general setting of quantum stochastic processes which may be both continuous (such as the Diosi-Penrose~\cite{Diosi_87_PLA}, CSL~\cite{Ghirardi_90_PRA,Bassi_03_PhyRep} and SUV~\cite{aritro2,aritro3,aritro1,aritroPhD} models) or discontinuous (such as the GRW model~\cite{Ghirardi_1986}). We shall use some basic results from Martingale theory in stochastic analysis~\cite{oksendal2003stochastic,revuz1999continuous} and define a physically viable collapse model as one which adheres to the probability processes being Martingales, so as to admit Born's rules exactly and avoid superluminal signalling~\cite{aritroFTL,Bassi2015,Gisin:1989sx,Bassi_03_PhyRep}. Concretely, we prove the following proposition:
\\\\
\noindent\textbf{\emph{Proposition:}}\,
\emph{A physically viable objective collapse theory, by itself, cannot dynamically result in thermodynamic equilibrium objectively (for single isolated systems or for all possible observables).}
\\\\
\noindent We prove the above proposition in two ways, \textbf{Proof I} is via contradiction and \textbf{Proof II} is via a more extended discussion which may be studied for further insight. 
\\\\
\noindent\textbf{Proof I} \emph{(by contradiction)}: Let $\mathbb{T}$ be an objective collapse theory, which collapses onto a particular collapse basis (such as the position basis for GRW) while also leading to systems achieving a thermal equilibrium state (where the density matrix is diagonal in the energy basis). Since the initial diagonal entries of the (not necessarily diagonal) density matrix in the energy basis necessarily change so as to attain thermodynamic equilibrium; the projections of the components of the energy basis density matrix, onto the collapse basis, will also change generically. Since the diagonal entries of the density matrix change in the collapse basis, $\mathbb{T}$ is not a physically viable objective collapse theory. \,$\Box$
\\\\
\noindent\textbf{Comments:} Without additional structure, a theory $\mathbb{T}$ allowing the diagonal entries of the density matrix to change generically in the collapse basis, cannot be a physically viable collapse model as Born's rules are not upheld in the dynamics. Further superluminal signalling is guaranteed in such a collapse model since deviations from Born's rules may be discerned by spatially separated parties~\cite{aritroFTL,Bassi2015,Gisin:1989sx}. If there is additional structure, which negates this change while also achieving thermal equilibrium, then $\mathbb{T}$ is not just a collapse model and would necessarily resemble the constructions in Sec.~\ref{Sec4:SUI}.
\\\\
\noindent\textbf{Proof II}: The above proposition is now proven by considering the following three statements:
\\\\
\noindent\textbf{(I)} \emph{In any physically viable objective collapse model, the diagonal elements of the density matrix in the collapse basis evolve as Martingale processes.}
\\\\
\noindent\textbf{Comments:} This is well known in the objective collapse literature~\cite{Pearle_89_PRA,Ghirardi_90_PRA,Bassi_03_PhyRep,aritro2,aritro3,aritroPhD}. Any physically viable objective collapse model must ensure that at long times, the steady state distributions correspond to Born's rules for ensembles. For any initial state of a system, $\ket{\psi(t=0)}$ undergoing objective collapse in the basis given by the complete set $\{\,\ket{e_i}\,\}_{i\,>0}$, the probability of obtaining the component $\ket{e_j}$ as the final result is given by Born's rules, $|\bra{e_j}\psi(0)\rangle|^2$, and must remain so at all times. Otherwise, spatially separated parties may discern the violation of Born's rules ~\cite{aritroFTL,Bassi2015,Gisin:1989sx} and send superluminal signals. Thus the probability processes for each component in the collapse bases, $z_j(t)=|\bra{e_j}\psi(t)\rangle|^2$, which are also the diagonal elements of the corresponding density operator ($\mathbb{E}[z_j]=\langle e_j|\hat{\rho}|e_j\rangle$) under stochastic averaging ($\,\mathbb{E}[\,..\,]\,$), must remain unchanged in any objective collapse model with causality. In other words, for all times $t>0$ and all its components, $j$, we have $\mathbb{E}[z_j(t)]=z_j(0)$, which implies $\mathbb{E}[ dz_j (t)]=0$, which finally implies that it is a pure (Ito) Martingale~\cite{oksendal2003stochastic,revuz1999continuous}, i.e. $\frac{\partial}{\partial t}\bra{j}\hat{\rho}\ket{j}=0$ ($\forall j, t$) or equivalently $dz_j(t)\propto dW_t$ in the continuous case. This holds true for all physically viable objective collapse models.
\\\\
\noindent\textbf{{(II)}} \emph{Martingale processes necessarily preserve the memory of its initial conditions. Thus, objective collapse models preserve the information of the initial conditions, as is also seen via Born's rules.}
\\\\
\noindent\textbf{Comments}: The first part of the above statement is a fact for any Martingale process (also see~\cite{oksendal2003stochastic,revuz1999continuous}). In our case, since the quantum probability process for each component is a Martingale, $dz_j(t)\propto dW_t$, expressed in the continuous Ito convention; we clearly have $\mathbb{E}[z_j(t)] = \mathbb{E}[z_j(0)]=z_j(0)$ since the initial conditions were definite. Hence Born's rules require the information of the initial state be (atleast partially) preserved and any objective collapse model must possess the Martingale property. 
\\\\
\noindent\textbf{{(III)}} \emph{A vector-valued stochastic process which induces a loss of the information of the initial state (such as a process leading to thermal equilibrium) cannot be a pure Martingale in all its components.}
\\\\
\noindent\textbf{Proof}: Let $\ket{\Phi(t)}\in \mathcal{H}$ follow a quantum stochastic process and its noise averaged statistical operator be $\hat{\rho}_\phi$. In the (separable) complete basis $\{\,\ket{e_i}\,\}_{i\,>0}\subset\mathcal{H}$, its probability components $z_i(t):=|\langle i\ket{\Phi(t)}|^2$ follow a stochastic process with some law (Ito): $d\mathbf{z}(t) = \mathbf{A}(\mathbf{z},t)dt + \mathbf{B}(\mathbf{z},t)d\mathbf{W}_t$, where $\mathbf{z}(t)\equiv (z_1(t),\, z_2(t),\, ..\,)$ is vector-valued. Its drift, $\mathbf{A}(\mathbf{z},t)$ and the total diffusion term, $\mathbf{B}(\mathbf{z},t)\, d\mathbf{W}_t$ are also vector-valued and in general, depend on the components of $\mathbf{z}(t)$ and time $t$. Let the initial conditions be $\mathbf{z}(0)$ and let the law of evolution result in (atleast partial) loss of information of the initial conditions in an ensemble of such processes. In other words, on average, the components of $\mathbf{z}(t)$ simply change in time and hence $\mathbb{E}[\mathbf{z}(t)]=\mathbf{z}(0)+\mathbb{E}[\mathbf{A}(\mathbf{z},t)]$ showing that it must atleast be a semi-Martingale with a drift. Thus $\mathbf{z}(t)$ cannot be a pure Martingale in all its components and thus each component $\mathbb{E}[z_i(t)]=\langle i|\hat{\rho}_\phi|i\rangle\neq z_i(0)$. Analogous arguments hold for discontinuous and jump processes.

Together, statements \textbf{{(I)}}, \textbf{{(II)}} and \textbf{{(III)}} necessarily imply that standard thermodynamic equilibrium is never dynamically achieved in any physically viable objective collapse theory (without further modifications). $\Box$
\\\\
\noindent\textbf{Comments}: The approach to thermal equilibrium ensures that the information of the initial state of the system is lost, such that only certain macroscopic quantities such as the statistical temperature and the mean energy are the only remaining characterization parameters; such processes cannot be pure Martingales in all its components. For any objective collapse model, the probability process in the collapse bases must be pure Martingales so as to adhere to Born's rules and avoid superluminal signalling, this implies that generically, objective collapse models do not possess the necessary and sufficient mathematical structure to admit an objective notion of thermalization or they allow superluminal signalling~\cite{aritroFTL}. Said differently, an approach to standard equilibrium thermodynamics requires that probability components in the energy basis cannot be pure Martingales, thus it cannot be a Martingale in any other basis either since the projections of this Martingale basis onto any other basis, such as the energy basis, will necessarily survive, which implies that thermal equilibrium is never achieved (solely due to objective collapse).

\textcolor{black}{We also note that in Ref.~\cite{Ag_DavidAlbert2021} numerical investigations have been carried out and they show that the GRW model does not lead to thermodynamic equilibrium. Further the use of collapse models which do not preserve the energy of the system is deemed problematic.  We note that so called dissipative CSL (dCSL) models have also been introduced which bound the unphysical energy violations of standard CSL by introducing a dissipative mechanism via a finite-temperature stochastic noise~\cite{Carlesso_2022,dCSL_2015}. The ontology of these models imply an open system and that the noise is a classical equilibrated environment (how or why the noise must equilibrate is not considered). The `thermalization' in these models refer to systems undergoing this dCSL mechanism to a canonical (open system) Gibbs steady state, with a temperature associated with the noise. \textcolor{black}{This steady state does not correspond to microcanonical equilibrium, valid for isolated systems and moreover, the average energy changes.} In contrast, in our article, by objective thermalization, we mean a dynamical approach to microcanonical equilibrium for an isolated single system. Further, the SUI model in Sec.~\ref{Sec4:SUI} using the SUV collapse \textcolor{black}{suppresses changes in the average energy} and yields Born's rules (for thermodynamic devices) exactly~\cite{aritro2,aritro3}. Summarizing, objective collapse models, at long times converge on information preserving, steady states corresponding to Born's rules, while in an objective thermalization model, long-time steady states are thermal equilibrium states, devoid of information of the initial state. These considerations further motivate our exploration of SUI models in Sec.~\ref{Sec4:SUI}.}
\bibliography{biblio_AM}
\clearpage \onecolumngrid 

\end{document}